\documentclass[preprint,amsmath,amssymb,aps,pre]{revtex4-1}

\usepackage{graphicx}
\usepackage{dcolumn}
\usepackage{bm}
\usepackage{mathrsfs}
\usepackage{hyperref}
\usepackage[utf8x]{inputenc}

\begin{document}


\title{Phase behavior of hard sheared cube family}

\author{Kaustav Chakraborty}
\author{Sumitava Kundu}%
\author{Avisek Das}
\email{mcsad@iacs.res.in}%
\affiliation{School of Chemical Sciences, Indian Association for the Cultivation of Science, Kolkata, INDIA }%


\begin{abstract}
A sheared cube is made out of a cube by giving a shear to the body in one direction keeping one of the faces fixed. We investigate here the thermodynamic phase behavior of a family of such regular hard sheared cubes, each of the members of the family having a distinct angle made by the faces with the perpendicular on the fixed face. Hard particle Monte Carlo (HPMC) has been performed with these anisotropic building blocks resulting entropy-driven self assembly. Thereby computational evidence of discrete plastic crystal phase has been found in crystal. The discrete plastic crystal phase is known to form through the spontaneous self-assembly of certain polyhedra. Throughout the entire solid regime particle orientations exhibit strong specific correlations before melting into a liquid, without any evidence of freely rotating plastic crystal at lower density solid. It has been thoroughly observed that geometrical attributes of the shapes don't determine any of the properties that designate this orientational disorder phase reported here. We also find that particle's rotational symmetric axes and one of the rotational symmetric axes of the unit cell of the crystal have a strong relationship in their alignment in space. These results, achieved with shapes having crystallographic point group symmetry, are investigated as being consistent with the phenomenology of discrete plastic crystal phase established in earlier works with hard particles having non-crystallographic point group symmetry. 

\end{abstract}

\pacs{Valid PACS appear here}
\maketitle

\section{\label{sec:Introduction}Introduction}
Self assembly is a process where building blocks spontaneously form ordered aggregates to give some structure \cite{Whitesides2002}. Studying the structure and its formation has always been a very important field of research \cite{Li2016}. Self assembled phases of crystals and liquid crystals are important for material design \cite{Ma2022, Rodarte2013, Glotzer2007, Lin2017}. If the constituents are not interacting with any kind of potential other than hard interaction, their self assembly is solely entropy-driven and they yield a diverse range of interesting self assembled structures of crystals and liquid crystals \cite{Onsager1949, Alder1957,  Berryman1983, Frenkel1984, Frenkel1985, Veerman1992a, McGrother1996, Bolhuis1997, GlotzerScience2012}. The simplest model of this kind involves crystallization of hard spheres into face-centered cubic or hexagonal closed pack crystal \cite{Alder1957, Frenkel1984, Thomas2018}. Computer simulations have been conducted to understand phase diagrams of hard spheres \cite{Alder1957, Frenkel1984}, hard spherocylinders \cite{Monson1978, Frenkel1990, Frenkel1997} to comprehend the phases achieved through entropy governed self assembly. Hard convex polyhedra serve as good models for nanoparticles. Utilizing nanoparticle self assembly for material designing purpose has been a principal aim of material science research for the past few decades \cite{Alivisatos1996, Mirkin1996, Mirkin2015, Mirkin2010, Mirkin2016}. The goal is to realize functional materials with desired properties  by designing shapes of the nanoparticle in modern experiments. Researchers have conducted controlled self assembly experiments with nanoparticles of various shapes like cubes, truncated cubes, octahedra, truncated octahedra, tetrahedra, superballs etc. Physicists started executing computer simulation with these faceted particles to determine the maximum crystalline packing and the crystal structures formed by them \cite{Choi2012, Torquato2009, Torquato2018}. Modern day computer simulations have greatly aided in determining equilibrium phases formed by hundreds of such faceted particles at different finite pressures.  As a result, detailed phase diagrams have been established with shapes such as the hard ellipsoid family \cite{Frenkel1985}, rectangular parallelepiped family \cite{Escobedo2008}, spherocylinder family \cite{Monson1978, Frenkel1990, Frenkel1997}, hard superballs \cite{Chaikin2008, Dijkstra2012, Torquato2010}, truncated octahedron \cite{Henzie2011, Barnard2005}, truncated cube family \cite{Dijkstra2015, Henzie2011, Evers2013, Escobedo2015, Escobedo2024}, rhombic dodecahedra \cite{Vutukuri2014, RhombicDodecahedra2010}, triangular prisms \cite{Dijkstra2017, Filion2024}, truncated tetrahedra \cite{Damasceno2012, GlotzerDigitalAlchemy2015}  etc. The discovery of dodecagonal quasicrystal through computer simulation of hard tetrahedra demonstrates that particle shape and entropy are capable of forming highly complex ordered structures \cite{Akbari2011, Akbari2009}. These research efforts have motivated the entire community to explore the relationship between the shape of the nanoparticle and the expected phase under desired density ranges.

If crystals are formed with anisotropic building blocks, disorder has been documented due to particle orientation, as the building blocks can be oriented differently in three-dimensional space \cite{Timmermans1961, UmangAgarwal2011, GlotzerScience2012, DijkstraDumbbells2008}. In the research of orientational phases realized in crystals formed by such hard interacting polyhedra researchers have found orientationally ordered crystals with single or multiple orientations \cite{GlotzerScience2012}. Additionally researchers have found crystalline phases where particles freely rotate over the ensemble confirming the existence of plastic crystal phases in entropic system \cite{Reynolds1975, Udovenko2008, Vdovichenko2015, DijkstraDumbbells2008, Agarwal2011, Escobedo2024}. This completely orientationally disorder phase was previously reported in the context of molecular solids \cite{Vdovichenko2015}. Surprisingly, a third type of orientational phase was discovered in the experiments with molecular crystals and computer simulations with some particular polyhedra where orientations of the constituents, despite being disordered,  were not random and showed strong correlation \cite{Goodwin2015, Goodwin2020, Goodwin2021, Abbas2022, Karas2019,  Kundu2024, KunduSymmetry2024}. Similar phase with preferred orientations was also reported in two-dimension \cite{Glotzer2D2019}. They were reported as ``discrete plastic crystal''  phase in the literature. Computer simulation of convex polyhedron also demonstrated existence orientational phase differing from freely rotating plastic crystal. Further studies of colloidal clathrate revealed discrete mobility among a finite number of specific orientations \cite{Lee2023}. These intriguing investigations posed serious questions about the coupling between translation and orientations of the building blocks in crystalline phases. However understanding this issue is easier with models using anisotropic polyhedra as they lack the complexities of chemical interactions. Our recent study with hard regular convex polyhedron systems reported this orientational phase achieved in cubic crystals with a set of newly developed detailed analysis of its properties, which incorporates a rigorous notion of orientational order and disorder. The system retained multiple orientations and a set of unique orientations throughout, whether in a fixed orientational state at high packing fractions or in more discretely mobile states at relatively lower packing fractions. We suggested that the system exhibited a long-range orientational correlation in the orientational space, where particles with each of the unique orientation achieved in the system, were nearly homogeneously distributed throughout the crystal \cite{Kundu2024}. Subsequently we reported the existence of a  ``local rule'' in terms of alignment of the highest order rotational symmetric axes of the particle with the rotational symmetric axes of a particular order of the crystal in this phase which maintained long-range orientational correlation and orientational disorder in unit cell level. This ``local rule'' was claimed to be the source of the long range correlation in orientational space. This special kind of alignments present in the discrete plastic crystal phase was reported to give a shape specific post priori justification of the finite number of distinct orientations in this phase \cite{KunduSymmetry2024}. 

In this article, we explore the phase diagram of hard sheared cube family where each member of the family has gradually increasing angles of shear relative to the perpendicular drawn on the fixed face of the cube. The primary focus of this study is to illustrate how positional and orientational characteristics of the particles within the attained phases evolve as the geometry gradually changes while preserving the point group symmetry of the particle. Moreover, we delve into the orientationally disordered phase here with all the signatures of maintaining strong correlation and examine how they behave as the geometry is adjusted. Given the establishment of an explicit local relation in our prior work, between the symmetry of the polyhedron and the symmetry of the crystal at which they self-assemble, we aim to further probe such relationships here with greater detail and a level of scrutiny not previously achievable with earlier systems. 

\section{\label{sec:Model and Merthod}Model and Method} 
\begin{figure*}
    \centering 
    \includegraphics[scale=0.8]{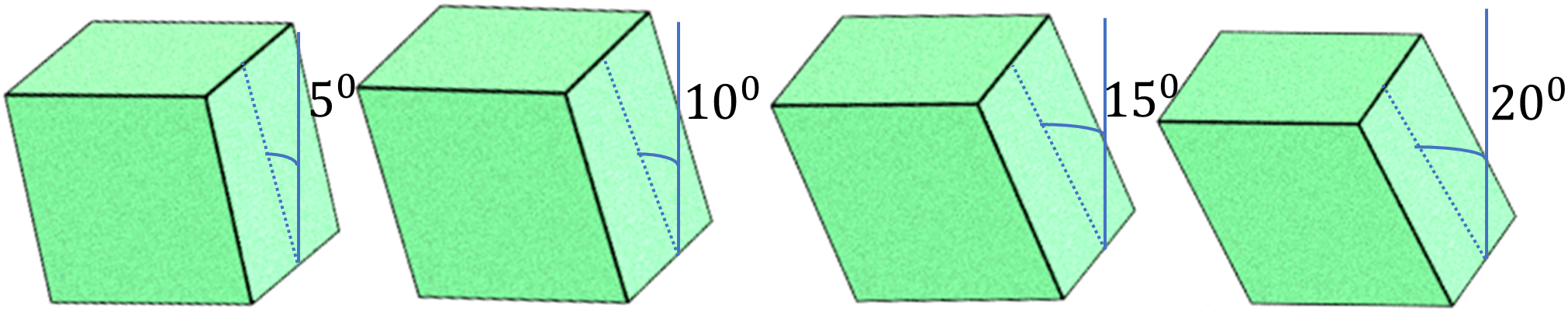}
    \caption{Sheared cubes with distinct angles}
    \label{fig:sheared_cubes_with_distinct_angles}
\end{figure*}

Sheared cubes with specific angles were made out of cubes by shifting the coordinates of the vertices of upper surface (where z-coordinates are positive, keeping the geometric center at the origin), while keeping the vertices of the lower (where z-coordinates are negative) surface fixed (Fig.\ref{fig:sheared_cubes_with_distinct_angles}). This transformation was performed under the constraint that edge length of all the edges of the polyhedron remained fixed. Subsequently, after this transformation, all of the vertices of the resulting sheared cube were recalculated to scale the volume of the body to unity. For varying degrees of shear, the vertices of the upper surface were horizontally translated  with the corresponding magnitude, ensuring the slant height was at the proper angle with the perpendicular drawn at the lower surface. Through this method, sheared cubes with distinct angles were achieved. 

\subsection{Simulation protocol}
We conducted Hard Particle Monte Carlo simulations using HOOMD-Blue toolkit \cite{Anderson2016}. According to the interaction model, when two particles overlap, the potential energy becomes infinite, while it remains zero if they do not overlap. The Metropolis algorithm was employed within HOOMD-Blue to handle this scenario. Initially, a system was prepared with very low packing fraction gaseous state (0.01 to 0.10 depending on the shape) to ensure no overlap in the system. Subsequently, the system was very slowly compressed to a packing fraction 0.55 to 0.60, depending on the shape. In the isochoric-isothermal(NVT) ensemble, long Monte Carlo sweeps were performed ($\sim$ 50 million), leading to equilibrium crystal phases whenever crystallization occurred. Following this, simulations were conducted in the  isobaric-isothermal(NPT) ensemble, gradually increasing the reduced pressure ($P^{*} = \beta  P v_{0}$, where $\beta = \left( k_{B} T \right) ^{-1}$ and $v_{0}$ is particle volume, which was set to 1.0 for every shape simulated) until reaching the highest packing fraction beyond which no noticeable change was observed. At each level of reduced pressure, sufficient equilibrium run was performed to ensure stable fluctuation of the box dimensions around the mean values. 
Subsequently, melting simulations were carried out in NPT ensemble, with successive decreases in total 20-25 values of reduced pressure until the system completely melted into isotropic liquid. At each packing fraction in the melting runs, were used to perform all analyses. 

\subsection{Orientational difference between two particles}
We used quaternion number systems to represent the orientations of particles \cite{Rowan2018}. The orientational difference between two regular convex polyhedra was measured as the minimum angle $\theta_{ij} = 2 \cos^{-1}[\Re(\mathcal{Q}^{\dagger}_i \mathcal{Q}_j)]$ between two quaternions, considering the proper point group of the particles ($\mathcal{Q}^{p}$). If the orientational difference is measured as zero between two particles, they are considered to be oriented in the same direction.

If the point group of the convex polyhedron has number of rotational symmetry operations that can be represented by a set of quaternions $\left\{ \mathcal{Q}^{p}_1, \mathcal{Q}^{p}_2 , ... , \mathcal{Q}^{p}_{n_{p}} \right\}$. Then the angle $\theta^{p}_{\gamma}$ is defined as $\theta^{p}_{\gamma} = 2 \cos^{-1}[\Re(\mathcal{Q}^{\dagger}_i \mathcal{Q}_j \mathcal{Q}^{p}_{\gamma})]$ for $ \gamma = 1, 2, ... , n_{p} $. The orientational difference between particles $i$ and $j$, denoted as $\theta_{ij}$ is defined as $\theta_{ij} = min \left\{ \theta^{p}_1, \theta^{p}_2 , ... , \theta^{p}_{n_{p}} \right\}$ \cite{Karas2019}. The histogram of these angles taking all pairs of particles in the system was calculated to understand the orientational ordering of the system. This was calculated by freud-toolkit \cite{freud2020}. 

\subsection{Detection of unique orientations}
In order to find out the absolute orientations of particles in three-dimensional orientation space, we selected three arbitrary reference particles from the system, ensuring significant variation in the angles between them. Subsequently, the orientational difference of each particle with these three references were plotted in three-dimensional space $\Theta_1$, $\Theta_2$ and $\Theta_3$. As each of the points in this space was a measure of absolute orientation with respect to these three chosen references, the number of clouds achieved in the three dimensional orientational space, is the number of unique orientations present in the system. 

An equivalent way of determining unique orientations of the system involved grouping each of the orientation by calculating mutual angles. Two particles $i$ and $j$ were said to have same orientations if the orientational difference $\theta_{ij}$ between them was less than a cutoff angle $\theta_{c}$, which was shape dependent and set well below the maximum allowed angle determined by the point group symmetry of the particle. This helped to categorize the particle into several disjoint sets, such that any two particles belonging to the same set had their pairwise orientational difference lower than the cutoff angle, while any two particle belonging to the different sets had their pairwise orientational difference higher than the cutoff angle. So, the number of disjoint sets $N_{\Omega}$ was the number of unique orientations present in the system. 

\subsection{Unit cell statistics based on particle orientations}
To classify unit cells according to the orientations of the particles belonging to it, each of the unit cells found from the systems were given unique indices. As the particles were already categorized according to the unique orientations they belong, they were assigned an unique index indicating the unique orientation, taking values in the range $[0, N_{\Omega}]$. For each unit cell we associated an array $\underline{\Lambda}$ of length $N_{\Omega}$. The $\Omega$'th element of the array is the sum of all particles in the unit cell with orientational index $\Omega$.
\begin{equation}
 \underline{\Lambda}(\Omega) = \sum_{i\in \mathbb{U}}\delta_{o_{i}\Omega}
\end{equation}
where, $\mathbb{U}$ is the set of indices of particles belonging to the unit cell and $o_i$ was the orientational index of $i$-th particle. One or more elements of the array could be zero, as all the unique orientations in the system might not be realized in every single unit cell. With this definition set, two unit cells a and b were said to be identical if and only if the following condition was true.
\begin{eqnarray}
	\underline{\Lambda}^a(\Omega) = \underline{\Lambda}^b(\Omega) \,\,\,\textrm{for all }\Omega.
\end{eqnarray}

Unique unit cells were counted and normalized distribution was calculated.

\subsection{Point group symmetries of the particle and crystal}
We analyzed the unit cells of crystal structure and the alignment of individual particle residing at a lattice site in the local frame of the unit cell. We represented the orientation of a unit cell by an orthogonal rotation matrix with respect to global laboratory frame $\mathcal{R}_{c}$ followed by the detection of rotational symmetry elements of the crystallographic point group $\Gamma^{c}$, within the unit cell's local frame. The axes of the rotational symmetry operations for the particle's point group $\Gamma^{p}$ were also determined in the particle's local frame.

We used the same classification of the rotational axes of the particle point group introduced by Kundu et al. and denoted the set of rotational symmetry axes of the constituent particle corresponding to the highest order as $\hat{\mathcal{A}}^{p}_{max}$, and all the rotational symmetry axes of the corresponding to the crystallographic point group as $\hat{\mathcal{A}}^{c}$ \cite{KunduSymmetry2024}. Using the global frame as the reference, we performed orthogonal transformations from the particle's local frame and its corresponding unit cell to the global frame, as described by the equations $\hat{\mathcal{S}}^{p}_{max, m, i} = \mathcal{R}_{i} \hat{\mathcal{A}}^{p}_{max, m} $ and  $\hat{\mathcal{S}}^{c}_{n} = \mathcal{R}_{c}  \hat{\mathcal{A}}^{c}_{n} $, where $\mathcal{R}_{i} $  represents the orthogonal rotation matrix defined in the global reference frame, corresponding to the unit quaternion $\mathcal{Q}_{i}$ of $i$-th particle. $\hat{\mathcal{A}}^{p}_{max, m}$ and $\hat{\mathcal{A}}^{c}_{n}$ are one of the elements of the set of axes corresponding to the highest order particle point group symmetry and all crystal rotational symmetry axes respectively extracted from corresponding to the point groups $\Gamma^p$ and $\Gamma^c$, where $m$ $\in$ [$1$, $\mathcal{N}^{p}$], $\mathcal{N}^{p}$ being the total number of $\hat{\mathcal{A}}^{p}_{max}$ axes and $n$ $\in$ [$1$, $\mathcal{N}^{c}$], $\mathcal{N}^{c}$ is the total number of $ \hat{\mathcal{A}}^{c}$ axes.

For each particle, using these transformed axes we could calculate pairwise angles. We calculated for each particle $i$, all angles $\alpha_{m, n}$, which is the angle between the unit vector  $\hat{\mathcal{S}}^{p}_{max, m}$ and all unit vectors $\hat{\mathcal{S}}^{c}$ axes. Then the minimum of all such angles was calculated according to eq.  

\begin{equation}
	\centering
	\alpha_{m,i} = \underset{n} {\text{\textit{min} }} \{\alpha_{m, n}\} = \underset{n} {\text{\textit{min} }} \{\cos^{-1}(\hat{\mathcal{S}}_{max, m, i}^{p} \cdot \hat{\mathcal{S}}_{n}^{c})\}
\end{equation}

Here, $\alpha_{m,i}$ represented the per-particle angle for m-th axis of the particle belonging to the set $\mathcal{S}^{p}_{max}$. In this article the data were presented as two-dimensional histogram of $\alpha_{m}$, for all particles with  $\mathcal{S}^{c}$, directly illustrating the minimum angles formed by each of the highest order rotational symmetry axes of the particle, $\mathcal{S}^{p}_{max, m}$ with any of the rotational symmetry axes of the corresponding unit cell of the crystal structure $\mathcal{S}^{c}$. 

This distribution served as an order parameter for distinguishing certain orientational phases. In a plastic crystal where orientational distribution is random, this distribution will exhibit a Gaussian nature. However, in the phases where the m-th rotational symmetry axis of the particle, corresponding to highest order $\hat{\mathcal{S}}_{max, m}^{p}$, was nearly parallel to any of the rotational axes of the crystallographic symmetry, $\hat{\mathcal{S}}^{c}$, the distribution showed a peak around $\sim$ 0$^{\circ}$. Conversely, when the axis  $\hat{\mathcal{S}}_{max, m}^{p}$ forms a finite angle ($\theta_{f}$) with any direction of crystal axis $\hat{\mathcal{S}}^{c}$, the distribution of $\alpha_{m}$ would sufficiently deviate from 0$^{\circ}$. 

Following the symmetry specific representation for a polyhedral shape of the general description of $\mathcal{S}^{p}_{max}$, $\mathcal{S}^{p}_{max} (C_{a})$ was introduced where $C_{a}$ represented a specific rotational operation, where $a$ is an integer denoting the number of folds in the rotational operation in the point group of the shape. As all the members of sheared cube shape family studied here have $D_{2h}$ point group the alignments of $\hat{\mathcal{S}}^{p}_{max} (C_{2})$ offered a full description of the orientation of the shape as $D_{2h}$ point group contains three mutually perpendicular $C_{2}$ axes in it, without any other proper rotational symmetric axis of different fold of rotation. 

\subsection{Transition matrix to detect discrete orientational hopping}
The transition matrix, $\underline{\underline{T}}$, is a $N_{\Omega}\times N_{\Omega}$ matrix characterizing the hopping of particles between different unique orientations along a particular trajectory. The number of transitions between two unique orientations with indices $w$ and $v$ for the particle $i$, $\mathcal{N}_i(w,v)$ is the number of times the particle toggled between these two particular orientations along the trajectory, keeping the order preserved, between two consecutive frames. $w=v$ indicates that particle did not change orientation. The elements of particles-averaged transition matrix is defined as follows.
\begin{eqnarray}
	\underline{\underline{T}}(w,v) = \frac{1}{N}\sum_{i=1}^{N}\frac{\mathcal{N}_i(w,v)}{(l-1)}
\end{eqnarray}
Here $l$ is the number of frames in the MC trajectory. The data were obtained over a trajectory length of $50\times10^{6}$ MC sweeps (MCS), and frames were saved at a frequency of $1\times10^{5}$ MCS.

\section{\label{sec:results}Results}
\subsection{Translational order}
\begin{figure*}
    \centering 
    \includegraphics[scale=0.8]{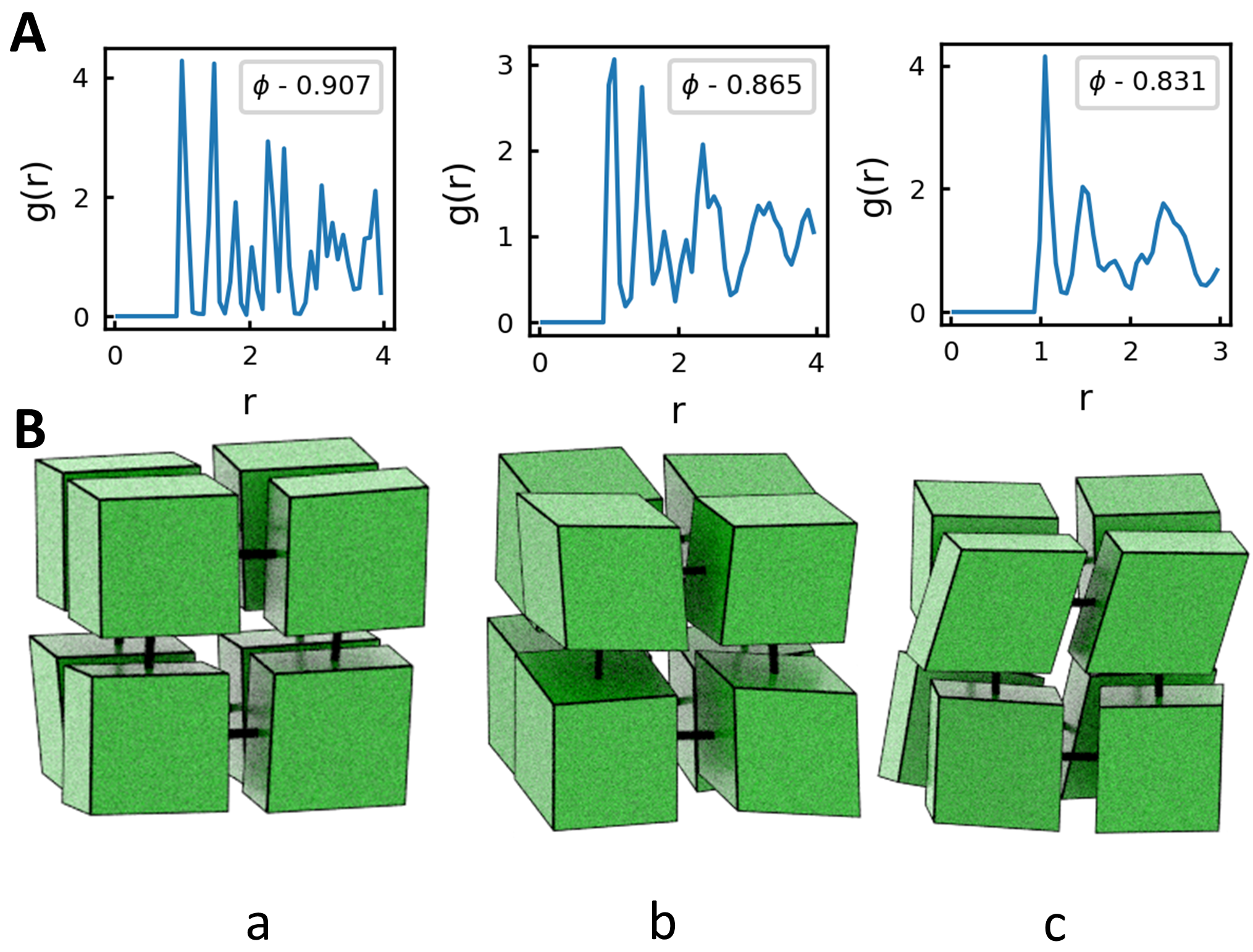}
    \caption{\textbf{Translational order in sheared cube family:} (A) Radial distribution function (RDF) analyses (B) Unit cell analyses. The subfigures indicate crystal like sharp peaks in RDF and simple cubic unit cells in sheared cube systems of $\delta = 5^{\circ}$ panel (a), $\delta = 10^{\circ}$ panel (b) and $\delta = 15^{\circ}$ panel (c).}
    \label{fig:translational_order_of_sheared_cubes}
\end{figure*}

Single-component systems consisting of sheared cubes with angles of $\delta = 5^{\circ}$, $\delta = 10^{\circ}$, and $\delta = 15^{\circ}$ were simulated to study the phase behavior with 8000 particles in each system. 
The positional ordering of the phase at highest packing fraction was analyzed by calculating radial distribution function (RDF) and determining the unit cell \cite{Kundu2024Algorithm}. At the highest pressure values, all systems composed of  $\delta = 5^{\circ}$, $\delta = 10^{\circ}$  and  $\delta = 15^{\circ}$ sheared cube exhibited simple cubic crystal structures (Fig.\ref{fig:translational_order_of_sheared_cubes}). The unit cell remained unchanged regardless of the shear angle, as previously defined. It was further observed that the highest achievable packing fractions achieved by NPT simulations after crystallizing around packing fraction 0.6, decreased with increasing shear angle and and the translational order deteriorated correspondingly. The unit cell remained simple cubic throughout the entire solid regions for each system, before they melted into isotropic liquid. 

\subsection{Orientational  order }
Orientations of the particle residing at lattice sites were represented by quaternions. For two particles with quaternions $\mathcal{Q}_i$ and  $\mathcal{Q}_j$, the orientational difference is measured by the angle between them defined as $\theta_{ij} = 2 \cos^{-1}[\Re(\mathcal{Q}^{\dagger}_i \mathcal{Q}_j)]$, after exercising proper consideration for the point group symmetry of the anisotropic body in question. For the bulk crystal achieved in the simulation, the average distribution of this orientational difference $\theta_{ij}$ captures the orientational phase to a good extent. To capture the global orientational phase behavior, in this study we measured all possible pairwise angles in the system.   

\begin{figure*}
    \centering 
    \includegraphics[scale=0.20]{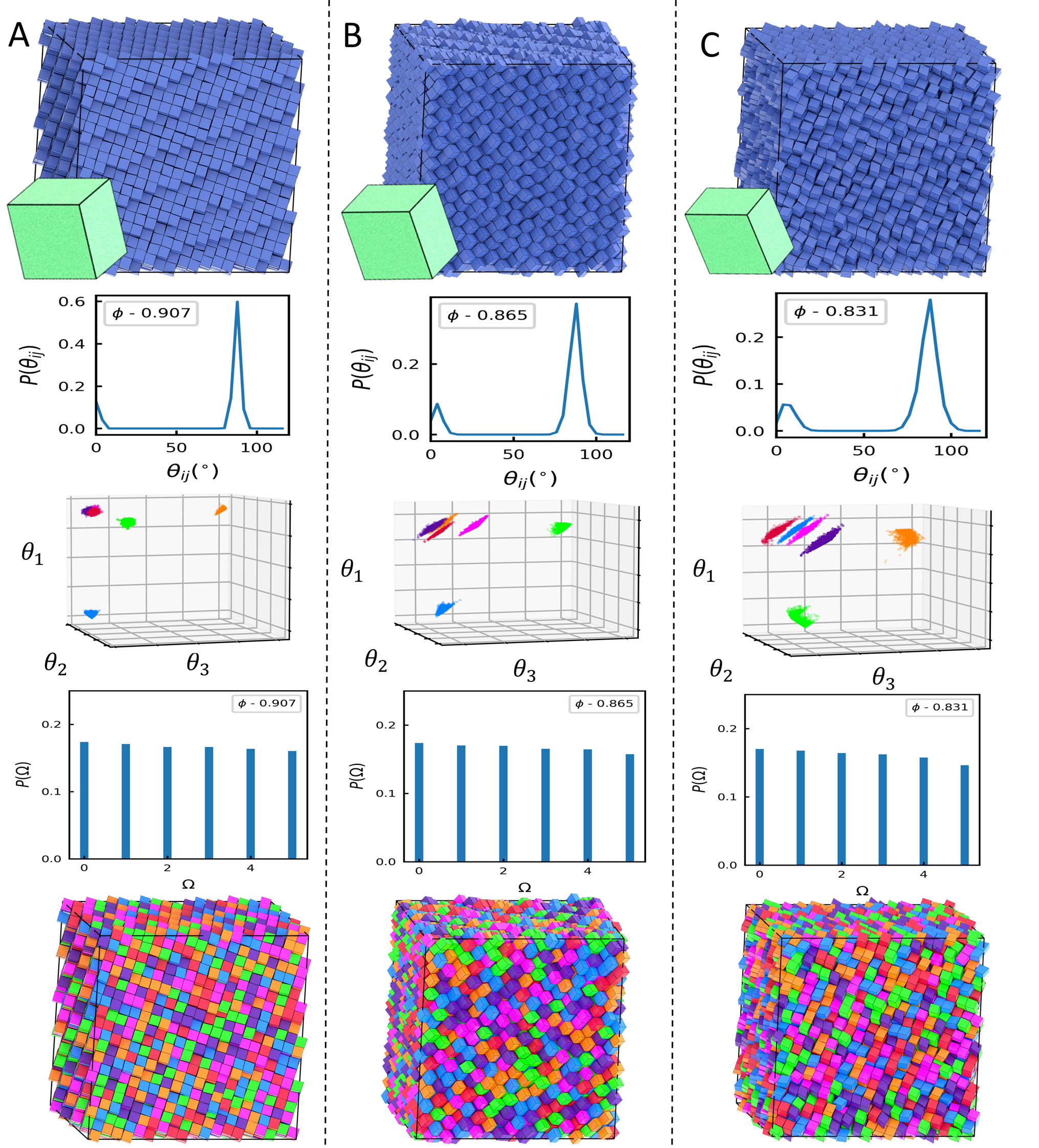}
    \caption{\textbf{Unique orientations and simulation snapshots at highest packing fraction in sheared cube family:} (A) $\delta$ = $5^{\circ}$ ($\phi \sim 0.907$) (B) $\delta$ = $10^{\circ}$ ($\phi \sim 0.865$) and (C) $\delta$ = $15^{\circ}$ ($\phi \sim 0.831$). Top portion of each panel represents the simulation snapshot at the densest packing. The distribution of pairwise angles calculated from the absolute orientations of the particles is shown in the second row of each panel. The third row depicts the absolute orientations of all particles in the simulation system in a three dimensional space of orientational differences from three reference orientations ($\Theta_{1}, \Theta_{2}, \Theta_{3}$ ).Particles with the same absolute orientations showed up as distinct clusters in these plots. Forth row shows population distributions among these orientations. Bottom figures show the snapshots of simulation systems with uniquely oriented particles rendered in a single color. All of the shapes have 6 unique orientations.}
    \label{fig:unique_orientations}
\end{figure*}

The top row of Fig.\ref{fig:unique_orientations}, under NPT, showcases snapshots of the systems at their highest achieved packing fractions. Upon visual inspection, assessing the orientational order necessitated further quantitative analysis. At highest packing fractions, the second row of Fig.\ref{fig:unique_orientations} illustrates the distribution of pairwise orientational differences between all pairs of particles from the system revealed two distinct peaks. These peaks stood apart by a significant height and maintained themselves $90^{\circ}$ apart. Altering the geometry of the bodies by increasing the shear angle seemed to have minimal effect on the distribution of pairwise orientational differences of all particles in the system, except for widening the peaks. For each of the shapes the peak corresponding to order with orientational difference $0^{\circ}$ was observed to be of significantly less height compared to that of the peak corresponding to disorder around $90^{\circ}$. Quantitative reproducibility of this non-random distribution between multiple independent runs of the same system size essentially confirmed the inter-particle orientational correlation in the system. Several well distinguished peaks in the orientation distributions were previously observed in the phase diagrams of water, as well as in simulations involving hard polyhedral particles and hard polygon shapes \cite{Goodwin2015, Karas2019, Glotzer2D2019}. Our recent study indicated that such distribution profile could serve as an order parameter for the system's many-body orientational states, allowing us to identify any distinct orientational patterns compared to a freely rotating completely random phase known as orientationally plastic crystal \cite{Kundu2024}.

But this pairwise angle distribution has its limitation as any straightforward correlation between this distribution and the orientational disorder present in the system was missing. The primary reason behind this was non uniqueness of a particle's orientation in three dimension from the only information that it makes orientational difference $90^{\circ}$ with some other. Kundu et al. \cite{Kundu2024} addressed this issue by applying a newly developed algorithm to detect the unique orientations present in the system. That approach is  independent of system size, removes any need for visual inspection and requires minimal intervention in the form of specification of an angle cutoff value to deal with statistical fluctuations. We calculated the orientational differences of a particle from three reference orientations ($\mathcal{Q}_{ref,1}$, $\mathcal{Q}_{ref,2}$, $\mathcal{Q}_{ref,3}$) from the system, defined as $\Theta_{1}$, $\Theta_{2}$, $\Theta_{3}$, by using the same procedure used in the calculation of $\theta_{ij}$. In order to calculate the angles denoted as $\Theta_{1}$, $\Theta_{2}$, $\Theta_{3}$ for all the particles in the system the following equation was used 
\begin{equation}
	\Theta_{ij} = 2 \cos^{-1}[\Re(\mathcal{Q}^{\dagger}_{ref,j} \mathcal{Q}_i \mathcal{Q}^p)]
\end{equation}
Where j = 1, 2, 3 and i was the particle index; i $\in$ $[1, N]$, N being the total number of particles in the system. $\mathcal{Q}^p$ represents the set of quaternions corresponding to the rotational symmetry operations in the point group of the particle, defined in the local frame. In the three-dimensional orientational space spanned by $\Theta_1$, $\Theta_2$, and $\Theta_3$, an absolute orientation appears as a unique point. Therefore, particles with the same absolute orientations would form clusters of points in a real system with statistical noise. The number of these clusters indicates the number of distinct orientations present in the system. Points within a cluster have $\theta_{ij}$ values less than a manually specified cutoff angle ($\theta_{c}$) set at least at the minima of the peak around $0^{\circ}$ of the distribution of pairwise orientational difference. This cutoff angle was dependent on the polyhedral shape. This method was applied to sheared cube shapes with different shear angles at their respective highest packing fractions achieved through self assembly. The detection of unique orientations in this manner revealed that each system was composed of a few discrete orientations with almost all particles. The number of unique orientations was six for all shapes, irrespective of shear angle (shown in six different colors in Fig.\ref{fig:unique_orientations} A, B, C, third row of each panel). The clouds of points in the orientational differences from three reference orientations supported these conclusions (Fig.\ref{fig:unique_orientations} A, B, C third row of each panel). The fraction of particles participating in each of the six unique orientations was calculated (Fig.\ref{fig:unique_orientations} A, B, C, fourth row of each panel) by performing a similar calculation of grouping the particles in the system into finite number of sets, where particles belonging to same set had their all mutual angles less than the cutoff angle $\theta_{c}$ as defined earlier with the minima of the sharp angular peak around $0^{\circ}$. This analysis clearly showed that all the unique orientations in each system were equally populated by the particles for each of the systems of sheared cubes with varying shear angles. Thus the particles were orientationally distributed among all the unique orientations realized in the system without showing any preference for any particular orientation. For all the  systems, particles with the same absolute angular dispositions are shown in same color, and the arrangements of the different orientations are depicted in bottom panels of Fig.\ref{fig:unique_orientations} A, B and C. All these behaviors were reproducible in multiple independent simulation for different system sizes. The unique orientations designated with distinct colors emerged in this phase describing orientations of the particles are randomly distributed all over across the system. Upon scrutiny of the distinct absolute orientations, the essence of disorder within this family of shapes revealed itself. Diversely oriented particles were dispersed haphazardly throughout the system, devoid of any discernible pattern. Distribution of pairwise orientational differences of particles belonging to different spatial distances from the origin confirmed the absence of any spatial dependence of the orientational distribution (See supplementary Fig.\ref{fig:distance_inv_supplementary} for details). All these characteristics of this orientationally disordered phase were also observed with $20^{\circ}$ sheared cube (See supplementary Fig.\ref{fig:deg20_si} for details). The pairwise orientational difference distribution exhibited even wider peaks, and the six clusters corresponding to the unique orientations became more dispersed in the pairwise orientational difference distribution relative to the three system references. 

All these features of orientational phase with finite number of distinct orientations achieved by computer simulation of sheared cube shape family with tunable shear angle, were reported with polyhedrons like EPD (Elongated Pentagonal Dipyramid) and ESG (Elongated Square Gyrobicupola) \cite{Kundu2024}. 

\begin{figure*}
    \centering 
    \includegraphics[scale=0.24]{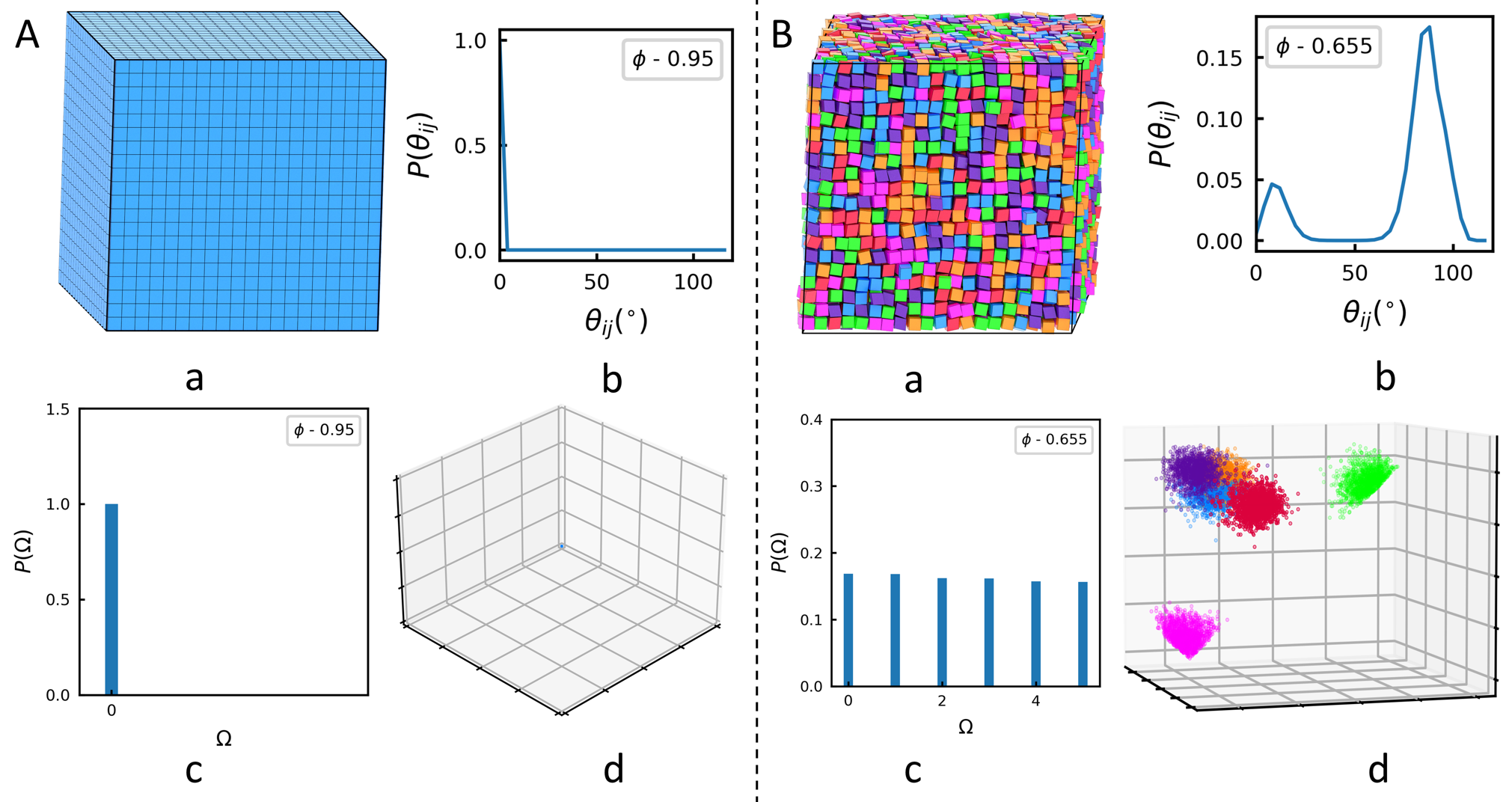}
    \caption{\textbf{Orientational disordered phase achieved for sheared cube $\delta$ = $10^{\circ}$ starting from high density ordered configuration.} A. Artificially generated orientationally ordered state with sheared cube $\delta$ = $10^{\circ}$ B. Correlated orientationally disordered phase achieved in NPT melting simulation with sheared cube $\delta$ = $10^{\circ}$. (a) System snapshot with particles belonging to a particular unique orientation rendered in a particular color. (b) Distribution of pairwise orientational difference of the particles. (c) Particle population distribution among different unique orientations in the system. (d) Clusters of orientational differences with respect to three system references plotted in 3D.}
	\label{fig:unpacking}
\end{figure*}

To further verify the robustness of this equilibrium phase, a simple cubic crystalline system was manually prepared at a very high packing fraction ($\phi \sim 0.95$) using cubes sheared by $10^{\circ}$, all aligned in a single orientation (Fig. \ref{fig:unpacking}). This configuration was never observed in self-assembly simulations, making it a suitable candidate for a different, potentially non-equilibrium initial condition. Moreover this configuration was prepared at very high packing fraction with sheared cube but could not have been achieved by any other shape previously reported in this context. An NPT simulation at $P^{*}=12 $, initiated from this artificially constructed state, resulted in a simple cubic crystal with six unique orientations, matching the orientational characteristics observed in spontaneous self-assembly simulations (Fig. \ref{fig:unique_orientations}). This phase was achieved after 120 million NPT MC steps. The equilibrated system retained all orientational characteristics associated with distinct orientations as described so far, aside minor statistical noise (Fig. \ref{fig:unpacking}). Consequently, the orientationally disordered phase could be attained both through self-assembly from an isotropic phase and from an artificially generated densely packed orientationally ordered state. This finding further confirmed the robustness and existence of this thermodynamic phase with hard interacting sheared cubes. 

\begin{figure*}
    \centering 
    \includegraphics[scale=0.24]{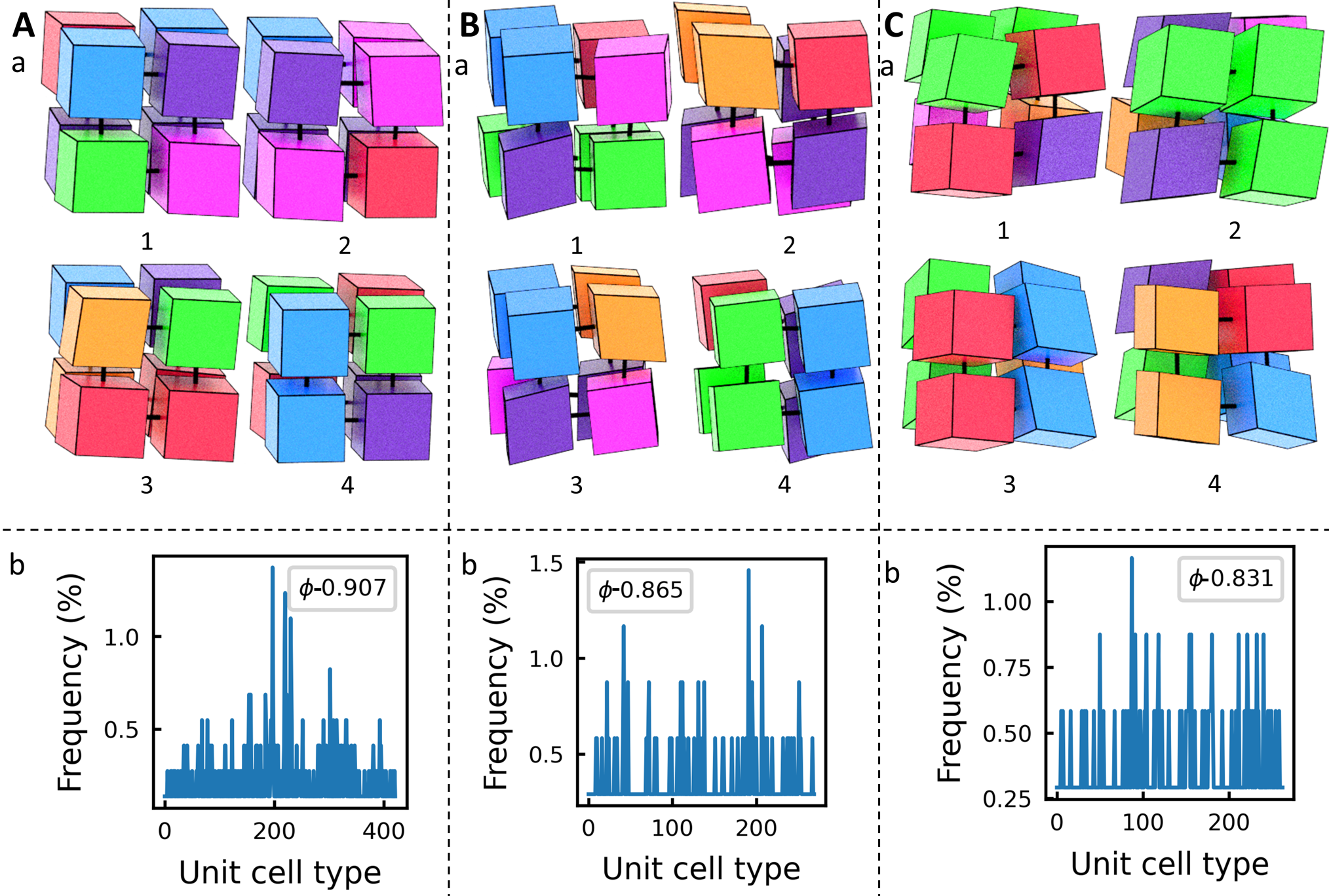}
    \caption{\textbf{Orientational analysis of unit cells at the maximum packing fractions. } Data for sheared cubes with $\delta$ = $5^{\circ}$, $10^{\circ}$ and $15^{\circ}$ are shown in panels A, B and C respectively. Particles are colored by their absolute orientations as in Fig. \ref{fig:unique_orientations}. Statistics of orientationally categorized unit cells are shown in the plots (for details, see the main text). All of the shapes were composed of large numbers of orientationally diverse unit cells. In all the systems there were no straightforward correlation between a particle position and its orientation in the many-body system. The nature of disorders of all these systems are therefore of similar kind.}
    \label{fig:uc_statistics}
\end{figure*}

Further delving into the quantitative comprehension of disorder necessitates a proper comparison with rigorously defined paradigms of orientationally ordered crystal and orientationally disordered crystal. In the realm of anisotropic particles, if the arrangement specifications within a single unit cell, upon crystal translations, can replicate the orientations of all particles in the bulk crystal, then the system attains orientationally ordered status.  It ensues from this definition that in an orientationally ordered perfect crystal of anisotropic particles, all unit cells exhibit translational and orientational equivalence. The orientational characteristics of a unit cell could be delineated by enumerating the quantities of particles possessing each distinct orientation (as outlined in the model and methods section). Kundu et al. \cite{Kundu2024} introduced this notion of disorder in terms of crystal unit cell was in this context of understanding non-random orientational disorder phase. With this analytical framework, it becomes feasible to gather statistical data on unit cell configurations from simulation outputs in finite and real systems, notwithstanding the presence of statistical noise. The absence of a dominant prevalence of any specific type of unit cell would indicate the presence of an orientationally disordered phase. The orientations of the members of the sheared cube family, relative to the corresponding simple cubic lattice, were demonstrated at the highest packing fractions achieved through self-assembly($\phi \sim 0.907$ for sheared cube with $\delta$ = $5^{\circ}$, $\phi \sim 0.865$ for sheared cube $\delta$ = $10^{\circ}$, $\phi \sim 0.831$ for sheared cube $\delta$ = $15^{\circ}$) in the form of unit cells comprising eight particles (Fig. \ref{fig:uc_statistics} a). This analysis revealed that none of the three particles connected via crystal translational vectors to a reference particle were orientationally ordered. However, each particle within the unit cell adopted one of the unique orientations present in the corresponding crystal. These unit cells exhibited rich orientational diversity, with no single configuration predominating within the complex many-body bulk systems (as depicted in Fig. \ref{fig:uc_statistics} a for each of the sheared cubes). Statistical analysis of the unit cell types confirmed that the systems consisted of a vast array of orientationally unique unit cells, with the prevalence of any single type being remarkably small, often only a few in number (illustrated in the bottom rows of each panel of Fig. \ref{fig:uc_statistics}). Consequently, at the unit cell level, the system appeared entirely random. Particle orientations exhibited no discernible relationship with the translational symmetry of the crystal. Furthermore, it was aligned with the visual insights into the orientational disorder present in the snapshots of the entire system (Fig. 3). This scenario precisely aligns with what was described in earlier literature regarding discrete plastic crystal phase with finite number of unique orientations, wherein the disorder is non-random at the level of individual particles but random at the level of unit cells \cite{Kundu2024}. This nature of unit cell level of disorder appeared to be similar for all characteristics for the variants of sheared cubes differing each other in geometry.

\subsection{Discrete rotational mobility of the disordered phase}

\begin{figure*}
    \centering 
    \includegraphics[scale=0.24]{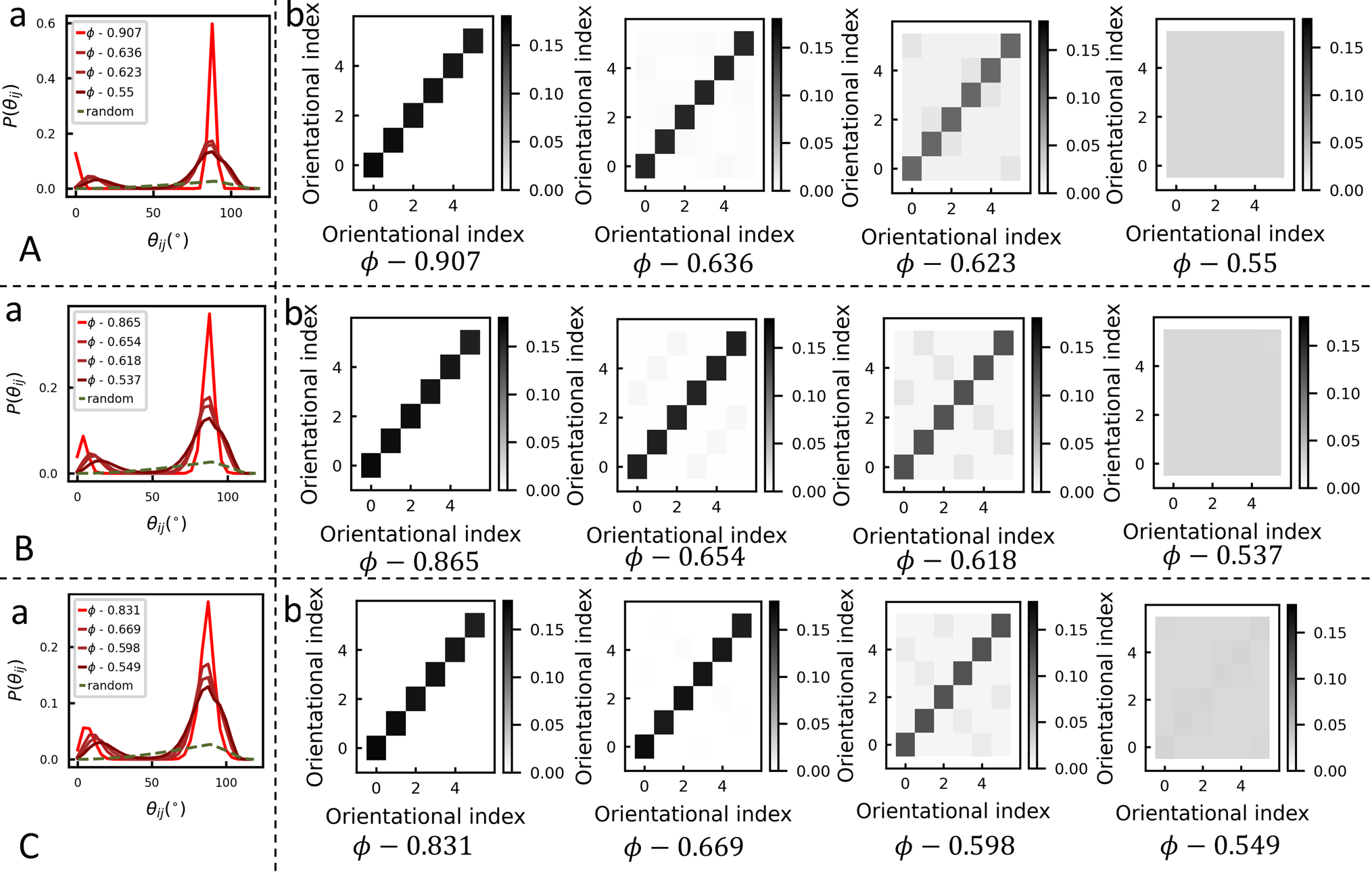}
    \caption{\textbf{Distribution of pairwise angle and signature of discrete orientational hopping with decreasing packing fractions are shown. } The distribution of pairwise angles calculated from the absolute orientations of the particles is shown in section a for sheared cubes with $\delta$ = $5^{\circ}$, $10^{\circ}$ and $15^{\circ}$ for decreasing packing fractions. The matrix illustrates the probability of particles transitioning from one unique orientation to another. At the densest packing, only the diagonal elements of the matrices are populated, indicating that no transitions occurred at the highest packing fractions. As the system begins to melt, the particles start toggling between the unique orientations.}
	\label{fig:lower_packing_fraction_behaviour}
\end{figure*}

In the high-pressure regions of the phase diagrams, the profiles of pairwise orientational differences between all pairs, represented by the distributions of $\theta_{ij}$, exhibited two distinct peaks for all three shapes (second row of each panel in Fig. \ref{fig:unique_orientations}). These peaks broadened as the pressure diminished, suggesting increased rotational mobility within lower density solids (as detailed for sheared cubes with $\delta$ = $5^{\circ}$, $10^{\circ}$, and $15^{\circ}$ respectively in Figure \ref{fig:lower_packing_fraction_behaviour} a). These characteristics persisted up to the packing fractions at which the solid phase remained stable for each shape before melting into isotropic liquid. For all the packing fractions spanning the solid regime in each of the sheared cubes the distribution of pairwise orientational difference deviated significantly from that distribution achieved by sampling random unit quaternion. The latter mimics the distribution of pairwise orientational difference of the same particle in a hypothetical phase where the particles are freely rotating over the ensemble, that's the orientationally completely disordered plastic phase \cite{Agarwal2011, GlotzerScience2012}. Consequently, none of the systems displayed any changes in their orientational distribution, contrary to the orientationally disordered phase observed in previously reported hard interacting systems with EPD and ESG \cite{Karas2019, Kundu2024}. As a result the sheared cube shape family did not exhibit any signature of freely rotating plastic crystal phase at low density regimes confirming absolute absence of any kind of solid-solid phase transitions involving the many-body behaviors of the particle orientations. These findings solidified the existence of robust inter-particle orientational correlations throughout the solid phase in systems comprising sheared cubes, regardless of the shearing angle - an observation further validated by consistent quantitative reproducibility across multiple independent simulation runs.

To gain a deeper understanding of how pressure affects orientational behavior at lower packing fractions, we examined the unique orientations obtained through a similar protocol and tracked any possible changes in orientation throughout the simulation trajectories. Data of angular distributions with respect to three reference orientations from system for each of the sheared cubes at their respective packing fractions at which pairwise orientational distributions were shown at Fig. \ref{fig:lower_packing_fraction_behaviour} i.e, packing fractions 0.907, 0.636, 0.623, 0.55 for $5^{\circ}$ sheared cube, 0.865, 0.654, 0.618, 0.537 for $10^{\circ}$ sheared cube and 0.831, 0.669, 0.598, 0.549 for sheared cube with $\delta$ = $15^{\circ}$ were studied (See supplementary Fig. \ref{fig:corr_dis_at_lower_pf}). We noticed quite similar kind of behavior compared to the previous studies, in the sense that  at each state point majority of the particles remained at six cluster of points in three dimensional orientation space for each of the sheared cubes. Thus, across the stability range, the solid phase still persisted the same set of discrete orientations appeared without changing the relative populations among them to any considerable level. The only difference in the distribution among six unique orientations of sheared cubes compared to the six unique orientations of ESG reported earlier at lower density solid, was the significant less statistical noise in terms of populations having extra orientations beyond the unique ones, as mentioned earlier \cite{Kundu2024}. This consistency aligned with the qualitative similarity observed in the profiles of the $\theta_{ij}$ distributions (outlined in  Fig. \ref{fig:lower_packing_fraction_behaviour} a for each of the sheared cubes). As a result, the complete dataset covering all members of this sheared cube shape family supported the claim that the presence of the orientationally disordered phase for these shapes aligned with prior findings of those with ESG, EPD and other polyhedra \cite{Kundu2024}. This phase was also marked by a stable count of unique orientations, equal population distribution, and consistent angular separations between them.

To examine the rotational mobilities of particles anchored at their designated lattice sites, we computed the transition matrix that delineates the average probabilities of transitioning from one orientation to another throughout a long equilibrium Monte Carlo trajectory (as outlined in the model and methods section). This approach mirrors the methodology previously employed to define the characteristics of this phase with ESG and EPD \cite{Kundu2024}. The results for all variants within the sheared cube family are presented in section b of Figure Fig. \ref{fig:lower_packing_fraction_behaviour}. The results validated the occurrence of discrete mobility between the selected orientations, ensuring that the overall partitioning remained constant throughout the simulation trajectory. At the highest packing fractions for each shape (i.e, $\phi \sim 907$ for sheared cube with $\delta$ = $5^{\circ}$, $\phi \sim 865$ for sheared cube with $\delta$ = $10^{\circ}$ and $\phi \sim 831$ for sheared cube with $\delta$ = $15^{\circ}$) there was an absence of orientational hopping, resulting in no off-diagonal entries in the transition matrix (as illustrated in the first column for each shape at section b). After significant melting of the system, no significant off-diagonal entries appeared in the transition matrix, indicating that the particles largely maintained their orientations, resulting in an orientationally frozen state throughout the system (as shown in the first column for each shape at Fig. \ref{fig:lower_packing_fraction_behaviour} b). The packing fractions up to which the off diagonal elements did not appear significantly, indicating no inter unique orientation rotational hopping, appeared to be higher with the increase in the shear angle of the sheared cube variants i.e, around $\phi \sim 0.636$ for sheared cube with $\delta$ = $5^{\circ}$, $\phi \sim 0.654$ for sheared cube with $\delta$ = $10^{\circ}$ and $\phi \sim 0.669$ for sheared cube  $\delta$ = $15^{\circ}$(as shown in the second column for each shape at Fig. \ref{fig:lower_packing_fraction_behaviour} b). As packing fractions further decreased (around $\phi \sim 0.623$ for sheared cube with $\delta$ = $5^{\circ}$, $\phi \sim 0.618$ for sheared cube with $\delta$ = $10^{\circ}$ and $\phi \sim 0.598$ for sheared cube with $\delta$ = $15^{\circ}$, off-diagonal elements began to appear, although particles still showed a strong preference for staying fixed in the preferred orientations they were, leading to dominant diagonal contributions (as shown in the third column for each shape at Fig. \ref{fig:lower_packing_fraction_behaviour} b). At sufficiently low packing fractions (around $\phi \sim 0.55$ for sheared cube with $\delta$ = $5^{\circ}$, $\phi \sim 0.537$ for sheared cube with $\delta$ = $10^{\circ}$ and $\phi \sim 0.549$ for sheared cube with $\delta$ = $15^{\circ}$, all chosen orientations were sampled by the same particles, and the off-diagonal elements achieved appreciable populations comparable to the diagonal elements (as shown in the fourth column for each shape at Fig. \ref{fig:lower_packing_fraction_behaviour} b) suggesting toggle within all the special orientations, following the same orientational correlations and exhibiting the qualitatively same kind of phase as observed at the highest packing fraction. This trend persisted, and just before transitioning to the isotropic liquid phase, maximum hopping was observed. The deviation of the distribution of pairwise orientational difference from that of the orientationally freely rotating plastic crystal phase indicated absence of any possibility of continuous transition from one unique orientation to the other leaving only possibility of discrete hopping. The orientational hopping data provided clear evidence of the discrete mobility with the hopping probability of the orientationally disordered state previously reported as ``discrete plastic crystal'' phase or ``discrete rotator phase''. This orientational hopping took place while particles consistently distributing themselves evenly among all unique orientations at all times (See supplementary Fig. \ref{fig:corr_dis_at_lower_pf} d for each sheared cubes).

\subsection{The local rule of correlation}
\begin{figure*}
    \centering 
    \includegraphics[scale=1.0]{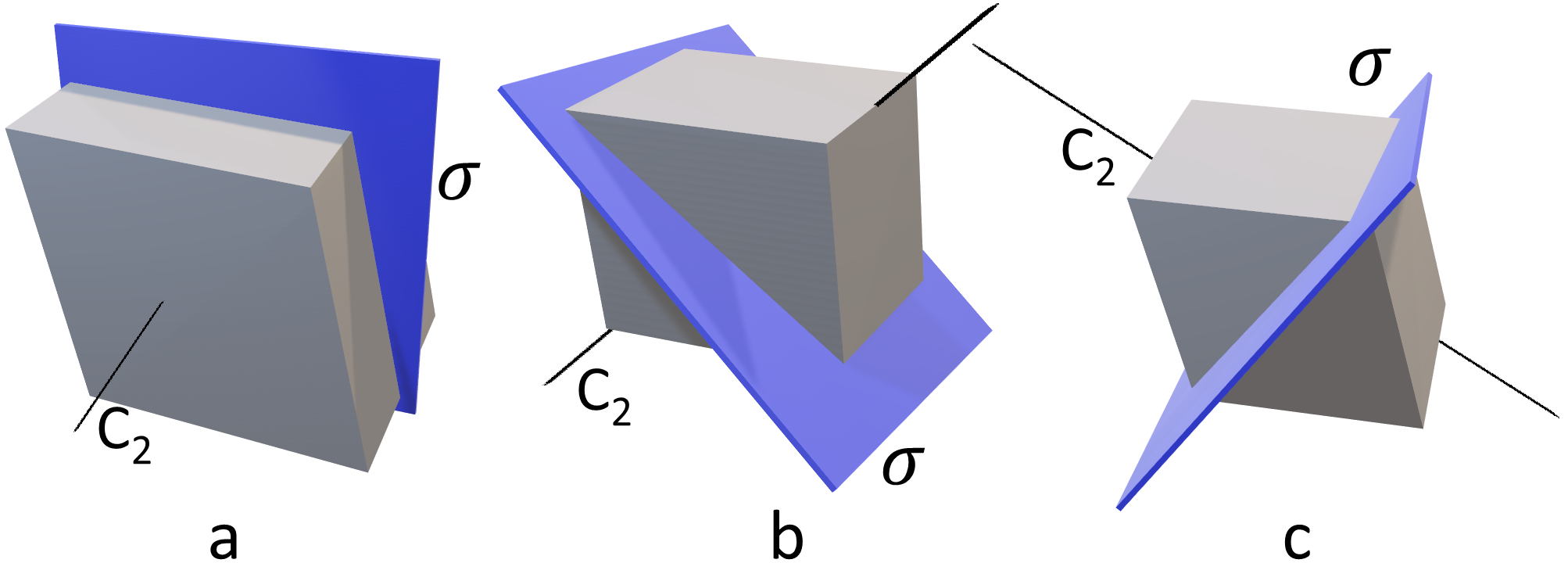}
    \caption{\textbf{Symmetry elements of point group of sheared cube shape family:} (a) $C_2$ axis passing through rhombus faces and $\sigma$ plane perpendicular to it. (b) $C_2$ axis passing through edges and $\sigma$ plane perpendicular to it. (c) $C_2$ axis passing through edges and $\sigma$ plane perpendicular to it. }
    \label{fig:sheared_cube_pg_symm}
\end{figure*}

Members of the sheared cube family possess rhombic dipyramidal $D_{2h}$ crystallographic point group symmetry, which belongs to orthorhombic crystal system. This symmetry features three twofold axes of rotation, each with three mirror planes perpendicular to each axis, as well as an inversion center (Fig. \ref{fig:sheared_cube_pg_symm}). The symmetry order of the sheared cube family is 8, which is significantly fewer than the $O_{h}$ symmetry group (belonging to cubic crystal system), which has a symmetry order 48. Therefore, this shape family clearly demonstrates that particles do not necessarily need to possess non-crystallographic point group symmetry to exhibit a correlated orientational disorder phase \cite{Kundu2024}. This finding contradicts earlier studies that suggested such phases were exclusively associated with shapes exhibiting non-crystallographic point group symmetry. Despite having crystallographic point group symmetry, the sheared cube family still demonstrates this disorder phase, suggesting that this phenomenon has broader applicability than previously thought. 

\begin{figure*}
    \centering 
    \includegraphics[scale=1.0]{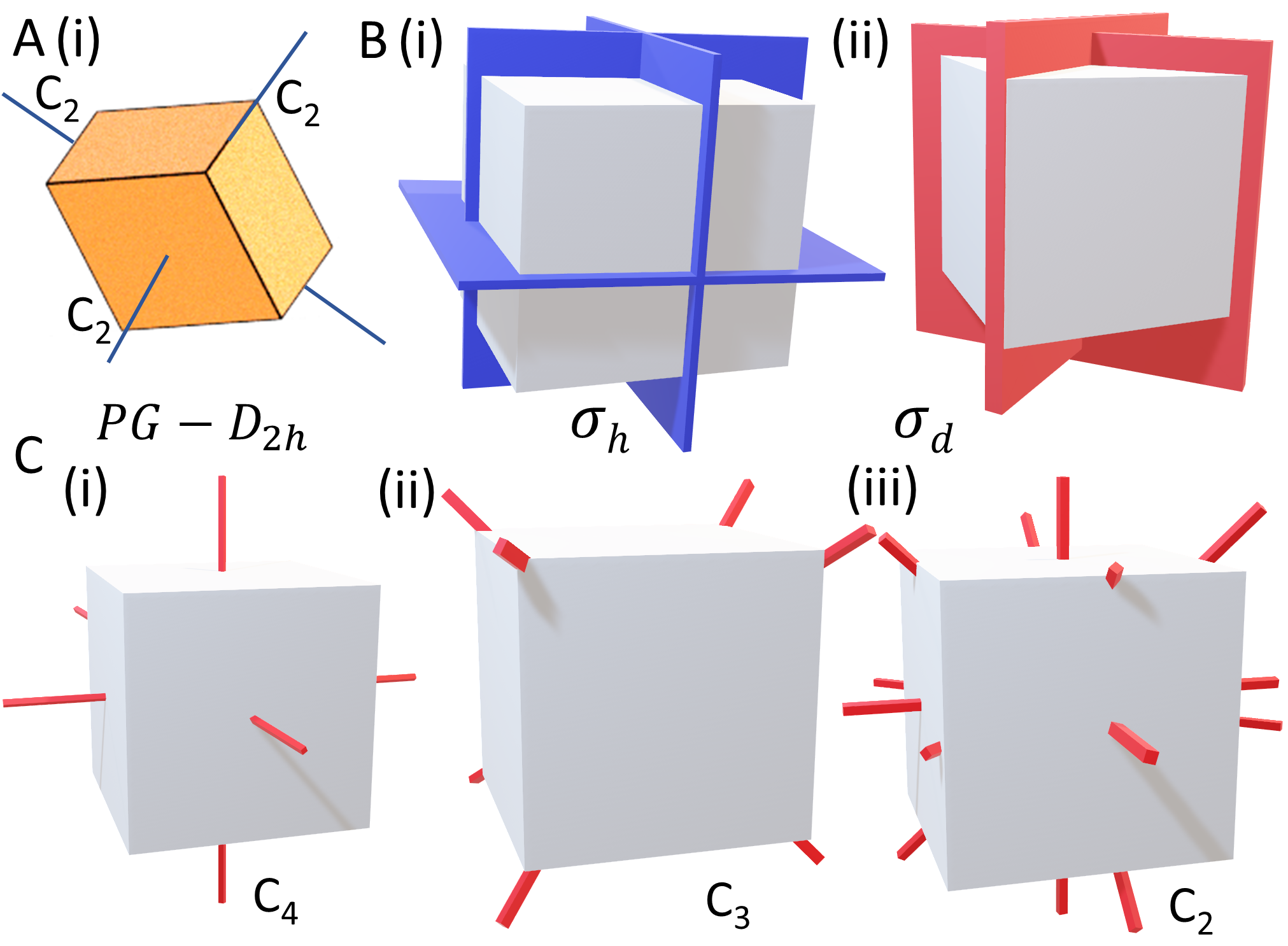}
    \caption{\textbf{Symmetry elements of a sheared cube with $\delta$ = $5^{\circ}$ and cubic crystal are shown.} rotational symmetry elements of the sheared Cube with $\delta$ = $5^{\circ}$ are displayed in the panel (A). The respective mirror planes and rotational symmetry elements of cubic crystal with $O_h$ symmetry are showcased in the panel (B) and (C) respectively.}
    \label{fig:sheared_cube_pg_symm_with_cube}
\end{figure*}

In line with previous studies, both the highest order rotational symmetry axes $\mathcal{S}^{p}_{max}$ of particle's point group and all rotational symmetric axes  $\mathcal{S}^{c}$ of crystallographic point group of the unit cell were considered to gain a better understanding of this phase (Fig. \ref{fig:sheared_cube_pg_symm_with_cube}). Unlike the shapes previously reported in the context of discrete plastic crystal phase, the rotational symmetry axes corresponding to the highest order separated by $90^{\circ}$, were not unique in the members of sheared cube family (Fig. \ref{fig:sheared_cube_pg_symm}). They feature three such $C_{2}$ axes: one passing through the opposite surfaces of the rhombus faces of the sheared cube body, and the other two through the diagonally opposite edges, all mutually perpendicular. As a result, the point group of the particle $\mathcal{S}^{p}_{max} (C_{2})$, is no longer remained unique, making the sheared cube family an ideal candidate for understanding the local rule previously suggested in the context of orientationally discrete plastic crystal phase. The variation in geometry while preserving the point group symmetry of the shape provided insight into the local rule within a family of bodies that differ solely in their geometry. 

The ensemble-averaged distribution of $\alpha$, as defined earlier, serves as the order parameter, highlighting a clear distinction between the discrete plastic crystal phase observed in the previously reported shapes and that of a freely rotating plastic crystal \cite{KunduSymmetry2024}. In this phase, the highest-order rotational symmetry axis $\mathcal{S}^{p}_{max}$ of particle's point group was reported to be nearly parallel to a particular rotationally symmetric axes $\mathcal{S}^{c}$ of crystallographic point group corresponding to the underlying crystal structures \cite{KunduSymmetry2024} . To deepen our understanding of this phase within the context of point group symmetry elements of polyhedral particles and their corresponding self-assembled crystal structures, research was conducted to determine which crystal symmetry axis, the $\mathcal{S}^{p}_{max}$ axes aligned with or were nearly parallel to (Fig. \ref{fig:sheared_cube_pg_symm_with_cube}). 
  
\begin{figure*}
    \centering 
    \includegraphics[scale=1.0]{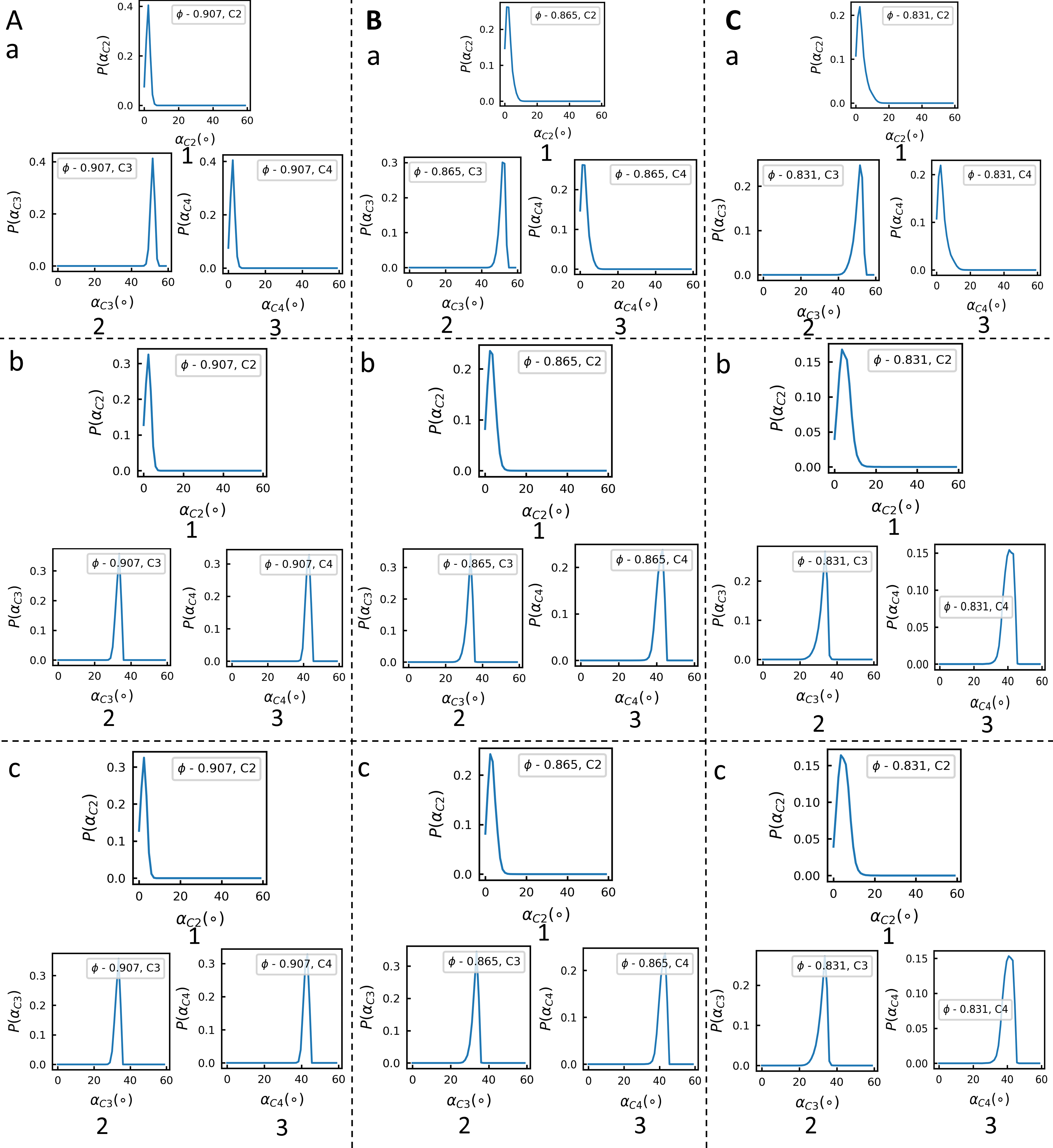}
    \caption{\textbf{Alignments of $\mathcal{S}^{p}_{max} (C_{2})$ axes of the particle with different rotational symmetric axes of the cubic unit cell are portrayed for three sheared angles :} (A) sheared cube with $\delta$ = $5^{\circ}$, (B) sheared cube $\delta$ = $10^{\circ}$, (C) sheared cube $\delta$ = $15^{\circ}$. Alignment of the $C_2$ axes of the particles with respect to the $C_2$, $C_3$ and $C_4$ axes of the respective cubic unit cells are shown in the subfigures (1), (2) and (3) indicating the directions of particle $C_2$ axes parallel to either $C_2$ or $C_4$ axes of the crystal.}
    \label{fig:axis_alignment}
\end{figure*}

The results, obtained from each member of the sheared cube family at their respective maximum achievable packing fractions ($\phi \sim 0.907$ for sheared cube with $\delta$ = $5^{\circ}$, $\phi \sim 0.865$ for sheared cube with $\delta$ = $10^{\circ}$, $\phi \sim 0.831$ for sheared cube) $\delta$ = $15^{\circ}$ are presented in Fig.\ref{fig:axis_alignment}(A,B and C). The first row for each of the shape (A for sheared cube with $\delta$ = $5^{\circ}$, B for sheared cube with $\delta$ = $10^{\circ}$ and C for sheared cube) $\delta$ = $15^{\circ}$ shows the distribution  of the angles made by the $\mathcal{S}^{p}_{max} (C_{2})$ axis passing through faces with the two-fold, three-fold and four-fold rotational symmetric axes of the cubic unit cell in their local frames Fig.\ref{fig:axis_alignment} a. For all variants of sheared cubes the probability distribution of $P(\alpha_{C_{n}})$ exhibited peaks at different $\alpha_{C_{n}}$ depending on $n$. At the distributions of angles made by face passing $\mathcal{S}^{p}_{max} (C_{2})$ with the two-fold axes and four-fold axes of the simple cubic unit cell, the peak appeared around $0^{\circ}$, indicating that the face passing $\mathcal{S}^{p}_{max} (C_{2})$ is nearly parallel to some of the two-fold and four-fold axes. In contrast, for the distribution of angles between the same axis and the three-fold axes of the simple cubic crystal, the peak appeared around $50^{\circ}$ indicating a distinct but non-parallel alignment with the three-fold axes of the underlying simple cubic crystal. 

Fig.\ref{fig:axis_alignment} b for each of the shapes are the distributions of the angles between one of the $\mathcal{S}^{p}_{max} (C_{2})$ axis passing through edges and the two-fold, three-fold and four-fold rotational symmetry axes of the cubic unit cell in their local frames respectively. In the distribution of angles between this edge passing $\mathcal{S}^{p}_{max} (C_{2})$ and the two-fold rotational symmetric axes of the crystal, a peak appeared around $0^{\circ}$, indicating a parallel alignment, while peaks at non-zero but distinct $\alpha_{C_{3}}$ and $\alpha_{C_{4}}$ values indicated a non-parallel but distinct alignment with the three-fold and four-fold rotational symmetric axes of the cubic unit cell. These signatures were consistent across all members of the sheared cube family. Distributions of the angles made by the other $\mathcal{S}^{p}_{max} (C_{2})$ axis passing through edges, with the same rotational symmetric axes of the simple cubic unit cell revealed a similar pattern to those of the previous edge passing $\mathcal{S}^{p}_{max} (C_{2})$ axis for each of the sheared cubes, indicating no significant difference in the angular distributions (Fig.\ref{fig:axis_alignment} c). This suggests that the face passing  $\mathcal{S}^{p}_{max} (C_{2})$ aligned in one way, while the edge passing  $\mathcal{S}^{p}_{max} (C_{2})$ axes aligned in a different way for each of the sheared cube within the local frame of the simple cubic unit cell. 

For each degenerate fold of rotational symmetric axes of the particle's point group, one common feature was that they aligned parallel with at least one of a particular fold rotational symmetric axes of the unit cell. This analysis implies that none of the rotational symmetry axes of the particle aligned parallel with any of the three-fold rotational symmetry axes of the cubic unit cell. As rotational symmetry axes of a particular fold present in the cubic unit cell are considered indistinguishable, and the four-fold rotational symmetry axes passing through the faces of the cubic unit cell are also the axes of two-fold symmetry rotations, the alignment of any specific rotational symmetry axis of the particle with $\mathcal{S}^{c} (C_{2})$ requires further analysis to determine whether it is aligned parallel with these common $\mathcal{S}^{c} (C_{2})$ - $\mathcal{S}^{c} (C_{4})$ axes, or with the $\mathcal{S}^{c} (C_{2})$ axes passing though the edges of the cubic unit cell, which don't share any common axis with any of $\mathcal{S}^{c} (C_{4})$.

\begin{figure*}
	\centering 
	\includegraphics[scale=0.24]{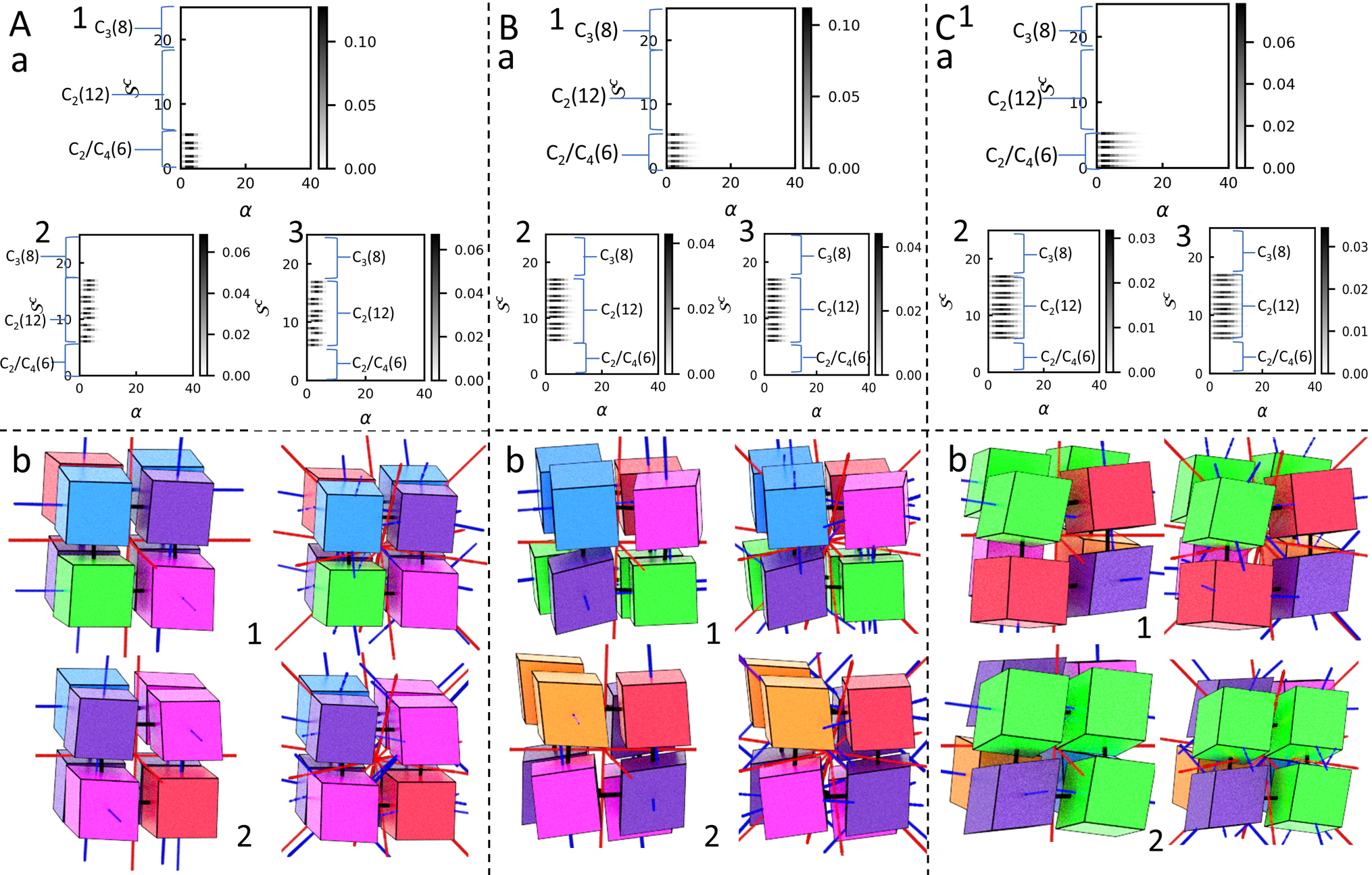}
	\caption{\textbf{Alignment of $\mathcal{S}^{p}_{max} (C_{2})$ axes of the particle with different crystallographic rotational symmetry axes $\mathcal{S}^{c}$.} (A) sheared cube with $\delta$ = $5^{\circ}$, (B) sheared cube with $\delta$ = $10^{\circ}$, (C) sheared cube with $\delta$ = $15^{\circ}$  Two dimensional distribution of angles $\alpha$ formed by $\mathcal{S}^{p}_{max} (C_{2})$ with the respective $\mathcal{S}^{c}$ axes (considered as distinguishable) are shown in the panel (a). The respective axes alignments of the particles at the level of unit cells are presented for visual confirmation as shown in the panel (b).}
	\label{fig:crystal_symmetry}
\end{figure*}

To obtain the statistical distribution of angles formed by the particle's highest order rotational symmetry axes corresponding to its point group with the rotational symmetry axes of the underlying unit cell's point group, all such angles were calculated and plotted in a histogram Fig. \ref{fig:crystal_symmetry} a, taking each of the rotational symmetric axis of the unit cell as distinguishable. Sections A, B, and C of Fig. \ref{fig:crystal_symmetry} a correspond to sheared cubes with $\delta$ = $5^{\circ}$ (at $\phi \sim 0.907$), $\delta$ = $10^{\circ}$ (at $\phi \sim 0.865$) and $\delta$ = $15^{\circ}$ (at $\phi \sim 0.831$), respectively. The upper section of each part shows the distribution of angles made by the $\mathcal{S}^{p}_{max} (C_{2})$ axes of the sheared cube family members passing through the rhombus faces. The lower section shows the distribution of angles made by the $\mathcal{S}^{p}_{max} (C_{2})$ axes of the sheared cube family members passing through the edges.
As the understanding of unique orientations in the system involves the alignment of the particle's point group rotational symmetry axes with those of the crystal (with statistical noise), only the average angles ($\alpha$ ) at which the particle's highest order symmetry axes make the minimum angle with the unit cell's rotational symmetry axes are plotted in the distribution. This approach provides a clearer depiction of the alignment behavior across different shear angles throughout the system.

In each plot (Fig. \ref{fig:crystal_symmetry} a), all 26 rotational symmetric axes of cube are aligned along the Y-axis, which simplifies the statistical analysis of the alignment between the highest order rotational symmetry axis of the particle's point group and the rotational symmetry axes of the point group of the unit cell of the underlying crystal. From the nonzero entries in the histogram, it becomes evident which rotational symmetry axis of the crystal is nearly parallel to the highest order symmetry axis of the particle. Statistically, it is observed that the $\mathcal{S}^{p}_{max} (C_{2})$ axes of the sheared cube family members, which pass through the surfaces of the opposite rhombus faces, align nearly parallel to the 6 common $C_{2}-C_{4}$ axes, i.e, $\mathcal{S}^{c} (C_{4})$ axes of the simple cubic crystal. Similarly, the $\mathcal{S}^{p}_{max} (C_{2})$ axes of the sheared cube family members that pass through the pairs of diagonally opposite edges align themselves nearly parallel to the twelve $\mathcal{S}^{c} (C_{2})$ axes of the point group of the simple cubic unit cell. The distribution of this angle defined for only with those $\mathcal{S}^{c}$ with which the corresponding $\mathcal{S}^{p}_{max} (C_{2})$ made the minimum angle were observed to be increasingly diffused away from $0^{\circ}$ with the increase of shear angle of the sheared cube particle. This analysis showing the distributions of the minimum angles formed by a particular $\mathcal{S}^{p}_{max} (C_{2})$ axis of the particles with $\mathcal{S}^{c}$ were found to be consistent with the distributions of angles formed by $\mathcal{S}^{p}_{max} (C_{2})$ axes of the particles with each of particular fold $\mathcal{S}^{c}$ axes taken to be indistinguishable. That's in the way that an axis aligned parallel with $\mathcal{S}^{c} (C_{4})$ would always be parallel with the face passing $\mathcal{S}^{c} (C_{2})$ as they share the same axis of rotation, but would always made a distinct finite non zero angle with $\mathcal{S}^{c} (C_{3})$. But an axis aligned parallel with $\mathcal{S}^{c} (C_{2})$ which do not share any common axis with $\mathcal{S}^{c} (C_{4})$ would never be parallel with the cubic unit cell's face passing $\mathcal{S}^{c} (C_{2})$ and with $\mathcal{S}^{c} (C_{3})$, but would always made a distinct finite non zero angle.

Fig. \ref{fig:crystal_symmetry} b includes images of two arbitrary unit cells from the system, with particles distinctly colored according to their respective unique orientations they are frozen at highest packing fractions. For both sample unit cells 1 and 2 of each sheared cube family member, the sample unit cell is shown twice side by side. The six $\mathcal{S}^{c}$ axes, all appearing identical and colored red, are shared by both the two-fold and four-fold rotational symmetric axes of the unit cell corresponding to the simple cubic crystal in one case.  In the other case, twelve $\mathcal{S}^{c}$ axes, also appearing identical and colored red, are shared by the two-fold rotational symmetric axes of the unit cell corresponding to the simple cubic crystal. The former passed through the faces of the cubic unit cell while the latter ones passed through the edges. 

In each of the particles from the specified random unit cells, all rotational symmetry axes of the point group symmetry are highlighted in blue. Consequently, the visual representation of the particles within the unit cells clearly delineates which of the $\mathcal{S}^{p}_{max}$ axes pass through specific sections of the polyhedron. Upon inspecting any particle in the first representation of each unit cell, it becomes evident that the $\mathcal{S}^{p}_{max} (C_{2})$ axes of the sheared cube family members, which pass through the surfaces of the opposite rhombus faces, are nearly parallel to some of the $\mathcal{S}^{c}$ axes. These are the axes shared by both the two-fold and four-fold symmetry axes of the point group of the unit cell corresponding to the simple cubic crystal. 

A similar analysis for another representation of the same unit cells reveals that the $\mathcal{S}^{p}_{max} (C_{2})$ axes of the point group of the sheared cube family members, which pass through the edges, align nearly parallel to one of the twelve $\mathcal{S}^{c} (C_{2})$ axes associated solely with the two-fold symmetry axes of the unit cell corresponding to the simple cubic crystal. This pattern holds across nearly all sampled unit cells from the systems of all members of the sheared cube family, with $\delta$ = $5^{\circ}$, $10^{\circ}$ and $15^{\circ}$. 

A noticeable difference between the six unique orientations of the hard sheared cube family and those of ESG system, previously reported \cite{Kundu2024, KunduSymmetry2024} in the context of this local rule of alignment of $\mathcal{S}^{p}_{max}$ with $\mathcal{S}^{c}$, is that in the sheared cube family, two distinct orientations are observed, differing by $90^{\circ}$, while the $\mathcal{S}^{p}_{max} (C_{2})$ axis passing through the rhombus face of the particle aligns with a particular $\mathcal{S}^{c} (C_{4})$. Visual inspection of second sample of a unit cell taken from the system with $\delta$ = $10^{\circ}$ (Fig. \ref{fig:crystal_symmetry} B.b.2) seared cube provides a clearer understanding. In the first representation of the second sampled unit cell, where face passing $\mathcal{S}^{p}_{max} (C_{2})$ of the particles are depicted in blue and cubic unit cell's $\mathcal{S}^{p}_{max} (C_{4})$  are depicted in red, it becomes evident that the particle with orange color (indicating one of the unique orientations achieved in the system) and the particle with purple color (indicating another unique orientation) both have their face passing $\mathcal{S}^{p}_{max} (C_{2})$ aligned parallel to a particular $\mathcal{S}^{c} (C_{4})$ axes in the unit cell. orientations of these two particles differed with each other by an angle $90^{\circ}$, as particles in different unique orientations always exhibit this characteristic in systems with hard sheared cubes. 

According to the point group of the particle, it has a mirror plane perpendicular to $\mathcal{S}^{p}_{max} (C_{2})$ axis. Kundu et al \cite{KunduSymmetry2024}. reported that presence of a horizontal mirror plane to the perpendicular to its $\mathcal{S}^{p}_{max}$, unique in that case, led to eight parallel alignments with $C_{3}$ axis passing through the 8 corners of the cubic unit cell, resulting in 4 unique orientations in the system. In this configuration, flipping the particle to the opposite alignment, where EPD's $\mathcal{S}^{p}_{max} (C_5)$ (which was parallel to $\mathcal{S}^{c} (C_{3})$) did not result in a different orientation of the particle. As reported in the context of ESG, the absence of any mirror plane perpendicular to $\mathcal{S}^{p}_{max}$, unique in that case, allowed 6 parallel alignments with the $C_{4}$ axis passing through the 6 faces of the cubic unit cell, leading to 6 unique orientations in the system. 

However, in the sheared cube system, here particles exhibited degenerate unique orientations, keeping the $\mathcal{S}^{p}_{max} (C_{2})$ axis passing through faces of the particle fixed, while interchanging the alignments of the other $\mathcal{S}^{p}_{max} (C_{2})$ axes passing through edges. Therefore, even though the presence of a mirror plane perpendicular to the $\mathcal{S}^{p}_{max} (C_{2})$ passing through faces of the particle, prevents the possibility of another orientation distinct from the one where the axis was aligned with a particular $\mathcal{S}^{c} (C_4)$ (such that aligning the same axis to the opposite face of the cubic unit cell would not create a different orientation) the system still exhibits 6 unique orientations, not 3. This is because, with every alignment of $\mathcal{S}^{p}_{max} (C_{2})$ axis passing through the faces of the particle along the $\mathcal{S}^{c} (C_4)$, the particle can realize two distinct orientations rather than just one.

This suggests that the alignment of the rotational symmetry axes of the polyhedron's point group with the rotational symmetry axes of the unit cell's point group - into which the particles self-assemble - is a symmetry-protected phenomenon, independent of the geometric variations between the different members of the sheared cube family.  Moreover, it is observed that when multiple highest order symmetry axes $\mathcal{S}^{p}_{max}$ exist in the particle's point group, all of the axes align along specific symmetry axes $\mathcal{S}^{c}$ of the corresponding unit cell's local frame. This highlights a systematic alignment within the crystalline structure.

This statistical behavior holds true across all members of the sheared cube family, confirming that the axis alignment is a symmetry-protected phenomenon, unaffected by the varying geometry of the shapes. Increasing the shear angle within this tunable shape family did not alter the distribution of mutual angles between the twofold symmetric axes in the point groups of the different shape variants. These axes aligned in the same manner, leading to the same number of unique orientations across all shapes in the family.

At lower packing fractions, during the melting of systems containing sheared cube variants of different shear angles within this family, the alignment of the axes remained largely unchanged, aside from minor variations caused by statistical noise present in lower-density solids (See supplementary Fig. \ref{fig:axis_alignment_at_lower_pf}). As particles in lower density states toggle between unique orientations, the axis alignment adjusts accordingly, preserving both the number of unique orientations  within the system and the distribution of pairwise orientational differences among all particles.

\subsection{Relation between all pairs of unique orientations}

\begin{figure*}
	\centering 
	\includegraphics[scale=1.3]{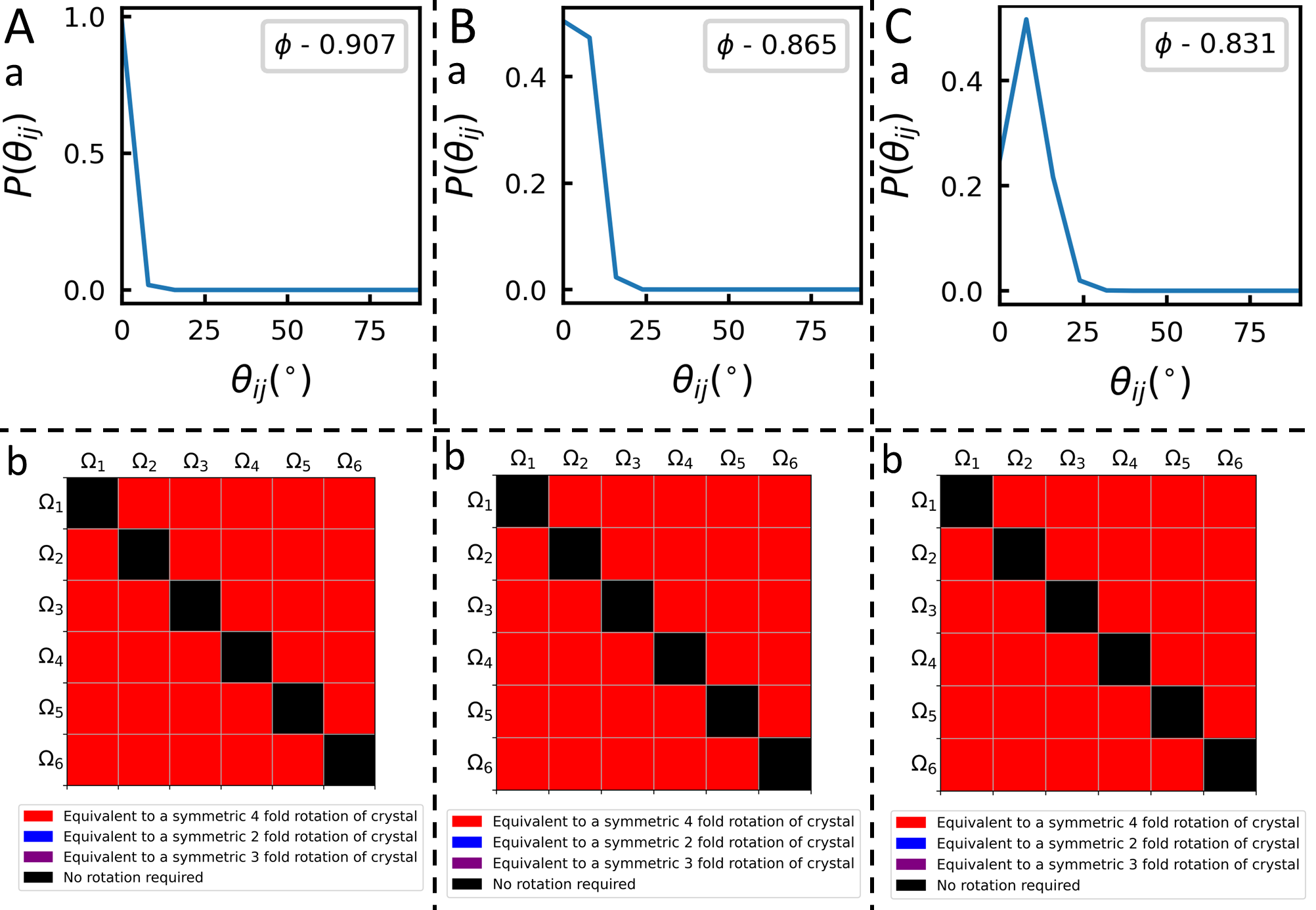}
    \caption{\textbf{Relation between all pairs of unique orientations:} (A) sheared cube with $\delta$ = $5^{\circ}$, (B) sheared cube with $\delta$ = $10^{\circ}$,  (C) sheared cube $\delta$ = $15^{\circ}$. Distribution of $\theta_{ij}$ made by the rotations relating one particle with the other, with crystallographic invariant rotational operation b. Order of rotational symmetric operation that relates all pairs of unique orientations realized in the bulk system.}
	\label{fig:relation_between_unique_orientations}
\end{figure*}

To understand the unique orientations within the system, we investigated how one unique orientation is related to the others. Since the three $\mathcal{S}^{p}_{max} (C_{2})$ axes are mutually perpendicular, they form an analogy of local orthogonal reference frame in the body. The orientations of these three axes characterize a specific unique orientation that the body assumes. Therefore, two different unique orientations correspond to two distinct orientations of an orthogonal set of axes and must be related by an orthogonal rotational operation.

We calculated the orientational differences between all pairs of particles within the system, as described in the methods section. These calculations were performed in the local frame of the cubic unit cell. The quaternion ($\mathcal{Q}$) for which for which the angle between $i$-th and $j$-th particle $\theta^{p}_{\gamma} = 2 \cos^{-1}[\Re(\mathcal{Q}^{\dagger}_i \mathcal{Q}_j \mathcal{Q}^{p}_{\gamma})]$ for $ \gamma = 1, 2, ... , n_{p} $  ($n_{p}$ is the number of quaternions representing the rotational symmetry operations of the body), becomes minimum, represents the rotational operation that relates the orientation of the $i$-th particle with that of the $j$-th in that frame. Furthermore, angles $\theta^{c}_{\epsilon} = 2 \cos^{-1}[\Re(\mathcal{Q}^{\dagger}_i \mathcal{Q}_j \mathcal{Q}^{c}_{\epsilon})]$ for $ \epsilon = 1, 2, ... , n_{c} $  ($n_{c}$ is the number of quaternion representing the rotational symmetry operations of the cubic unit cell)  were calculated. Thus, $\theta^{c}_{ij}$ is defined as $\theta^{c}_{ij} = min \left\{ \theta^{p}_1, \theta^{p}_2 , ... , \theta^{p}_{n_{c}} \right\}$, where $\theta^{c}_{ij}$ represents the minimum quaternion angle formed by the rotational operation connecting the orientations of the pair of particles, relative to the symmetric rotations of the cubic unit cell. A histogram of these angles, computed for all particle pairs in the system, was then constructed to understand how rotations relating the orientations of any two particles, belonging to either the same or different unique orientations, are related to the symmetric rotations of the underlying crystal structure. 

Fig. \ref{fig:relation_between_unique_orientations} a shows the distribution of $\theta^{c}_{ij}$ in the sheared cube family with $\delta$ = $5^{\circ}$, $10^{\circ}$ and sheared cubes $\delta$ = $15^{\circ}$ respectively. Each distribution exhibits a peak around $0^{\circ}$ $\theta^{c}_{ij}$, with the width of the peak increasing as the shear angle of the particle increases. The peak around $0^{\circ}$ indicates that the orthogonal rotational operations relating the orientation of one particle to that of the another - whether the particles belong to the same unique orientation or different ones - are predominantly symmetric rotational operations of the underlying cubic crystal. The increasing dispersion of the particle orientations with increasing shear(shown in Fig. \ref{fig:unique_orientations}) is responsible for the widening of the distribution of Fig. \ref{fig:relation_between_unique_orientations} with increasing shear angle of the particle. This relationship between the rotations connecting the orientations of all particle pairs and the symmetric rotations of the crystal structure offers deeper insights into the relationships between the unique orientations realized in the system. 

To further understand which crystallographic symmetric rotations matched with the rotations connecting two distinct unique orientations in the system, we did this calculation with quaternions representing the unique orientations of the bulk system system, this time considering the order of the rotational symmetry operation of the crystal. Fig. \ref{fig:relation_between_unique_orientations} b shows that all the rotations in the local frame of the unit cell, relating two different unique orientations realized in the bulk system, are equivalent to some foul-fold symmetric rotations (shown in red) of the underlying unit cell, without any presence of other symmetric rotations of the crystal. This was observed for all the non-diagonal elements of the matrix relating two different unique orientations ($\Omega_{i} \to \Omega_{j}$ and $\Omega_{j} \to \Omega_{i}$), while for diagonal elements, representing the crystallographic rotational symmetry operations relating a unique orientation to itself, the elements are shown in black, as no non-trivial rotational symmetric operation (other than identity operation) was required. 

This observation, for this shape family having rhombic dipyramidal $D_{2h}$ crystallographic point group across different shape variants with varying geometries, strongly suggests that the unique orientations realized in the system are governed by symmetry, with minimal influence from geometry.

\section{\label{sec:discussion}DISCUSSION}

We present evidence of a discrete plastic crystal phase characterized by strong orientational correlation among particles, despite unit cell-level disorder, in a family of regular convex polyhedra with tunable geometries. This phase emerges from a simple body composed of cubes that have been sheared in one direction, giving rise to a rich phenomenology of correlated orientational disorder within the self-assembled crystalline structure. In this phase, all translationally connected particles displayed orientational disorder, a phenomenon that manifests at the unit cell level within the crystal. This orientational disorder adheres to specific correlation laws, maintaining a non-obvious broken rotational symmetry state in simple cubic crystal structure. 

These findings highlight that this type of novel orientational disorder phase can occur even in primitive Bravis lattice like simple cubic in cubic lattice system, which observed this phase only in centered Bravais lattices with hard-interacting regular polyhedra \cite{Kundu2024}. This further reinforces the robustness of the orientational disorder phase, with strong correlations in orientational space, across different translationally symmetry-broken phases in three dimensions. Consequently, the potential existence of this phase in lattice systems other than cubic should be seriously considered in future studies. Moreover, the hard-interacting sheared cube family expands the scope of discrete plastic crystal phases, showing that hard particles with crystallographic point groups can also exhibit an orientationally disordered thermodynamic phase in cubic crystals. Thus, this shape family, along with other shapes exhibiting this phase, raises important questions. Even within the simple cubic crystal, the particles exhibited all the characteristics of discrete rotations among unique orientations in a correlated fashion, maintaining conserved quantities - such as the fixed number of unique orientations, equipartition of population among them, and the distribution of pairwise orientational differences-intact throughout the solid range of the phase. Notably, none of the systems exhibited any existence of freely rotating plastic crystal before they melted into isotropic liquid. The absence of a plastic crystal phase in lower-density solids across the entire shape family raises further questions: what determines the existence of a freely rotating plastic crystal in hard particle self-assembly, and does the crystal structure play a role in this determination? The lack of a solid-solid transition in these systems suggests a strong coupling between the translational and orientational degrees of freedom of the particles. The characteristics of the discrete plastic crystal phase, with strong orientational correlation, were found to be identical to those observed in systems with solid-solid phase transitions in their phase diagrams, indicating the broader existence and robustness of this phase in entropy-driven systems.  

This study of the symmetry-preserving tunable shape family also provides substantial evidence that this phase is symmetry-governed. Previous studies of different shapes with non-crystallographic point group symmetry, such as EPD and ESG, showed that the number of unique orientations varied with different point groups ($D_{5h}$ for EPD and $D_{4d}$ for ESG). The study of the hard sheared cube family revealed that the number of unique orientations depends solely on the point group of the particle, with other shape attributes playing no role. While the number of unique orientations in the sheared cube phase was found to match exactly with that of ESG, there were differences in the distribution of pairwise orientational differences. The data further indicate that the distribution of pairwise orientational differences, together with the fixed number of unique orientations in three-dimensional orientational space and the equal distribution of population among them, collectively characterizes this correlated orientational disorder phase. Achieving the same number of unique orientations in three dimensions while maintaining long-range orientational correlations in different ways, depending on the point group symmetry of the particle, underscores the significance of the particle's point group symmetry in governing the phase.

Building on the earlier study's observation of correlated orientational disorder, which established a one-to-one relationship between the alignment of the particle's highest-order symmetry axis and the crystal's symmetry axes with the unique orientations observed in the system, this relationship was confirmed in the current study.  Specifically, the $C_{2}$ axes of the sheared cube family members (all exhibiting $D_{2h}$ point group symmetry) that pass through the surfaces of the rhombus faces align parallel to the $\mathcal{S}^{c} (C_4)$ (six common $C_{2}-C_{4}$ axes) of the crystal. At the same time, the other $C_{2}$ axes within the particle, which are perpendicular to the first set, align with the 12 $C_{2}$ axes of the crystal. This finding confirms the validity of the ``local rule'' proposed in the earlier studies \cite{KunduSymmetry2024}, suggested the alignment of the highest-order rotational symmetry axis of the particle's point group with a particular rotational symmetry axis of the underlying unit cell governs the system's unique orientations. Importantly, this study extends the applicability of the ``local rule'' to particles with multiple highest-order symmetry axes in their point group. The rule now encompasses the alignment of all highest-order rotational symmetry axes of the particle with those of the crystal, depending on the particle's point group symmetry.

Unlike shapes with non-crystallographic point groups previously reported \cite{Kundu2024}, the distribution of pairwise orientational differences for all members of the sheared cube family showed distinct peaks at approximately $0^{\circ}$ and $90^{\circ}$. These angles correspond to the rotational symmetries of the cubic crystal structure, suggesting the possibility that two distinct unique orientations are related by a rotation about an axis at a crystallographically symmetric angle. This study further demonstrated that the rotational operations connecting every pair of distinct unique orientations in the local frame of the cubic unit cell are, in fact, the symmetric rotations of the cubic unit cell. Moreover, all such rotations belonged to a specific order of symmetry operations. 

This observation implies that the unique orientations observed in systems with particles belonging to a crystallographic point group can provide deeper insights into the role of symmetry in correlated orientational disorder phases. Specifically, it suggests that symmetry governs the selection of a particular set of crystallographic rotational symmetry operations that define the unique orientations in the system. This finding opens avenues for future investigations into the interplay between symmetry and correlated orientational disorder in hard particle systems.

\begin{acknowledgments}
KC thanks IACS for financial support. SK acknowledges financial support from DST-INSPIRE Fellowship (IVR No.\,201800024677). AD thanks DST-SERB Ramanujan Fellowship (SB/S2/RJN-129/2016) and IACS start-up grant. Computational resources were provided by IACS HPC cluster and partial use of equipment procured under SERB CORE Grant No.\,CRG/2019/006418.  
\end{acknowledgments}

\bibliography{sheared_cube}

\newpage
\section{Supplementary Information}
\begin{figure*}[ht]
	\centering 
	\includegraphics[scale=0.24]{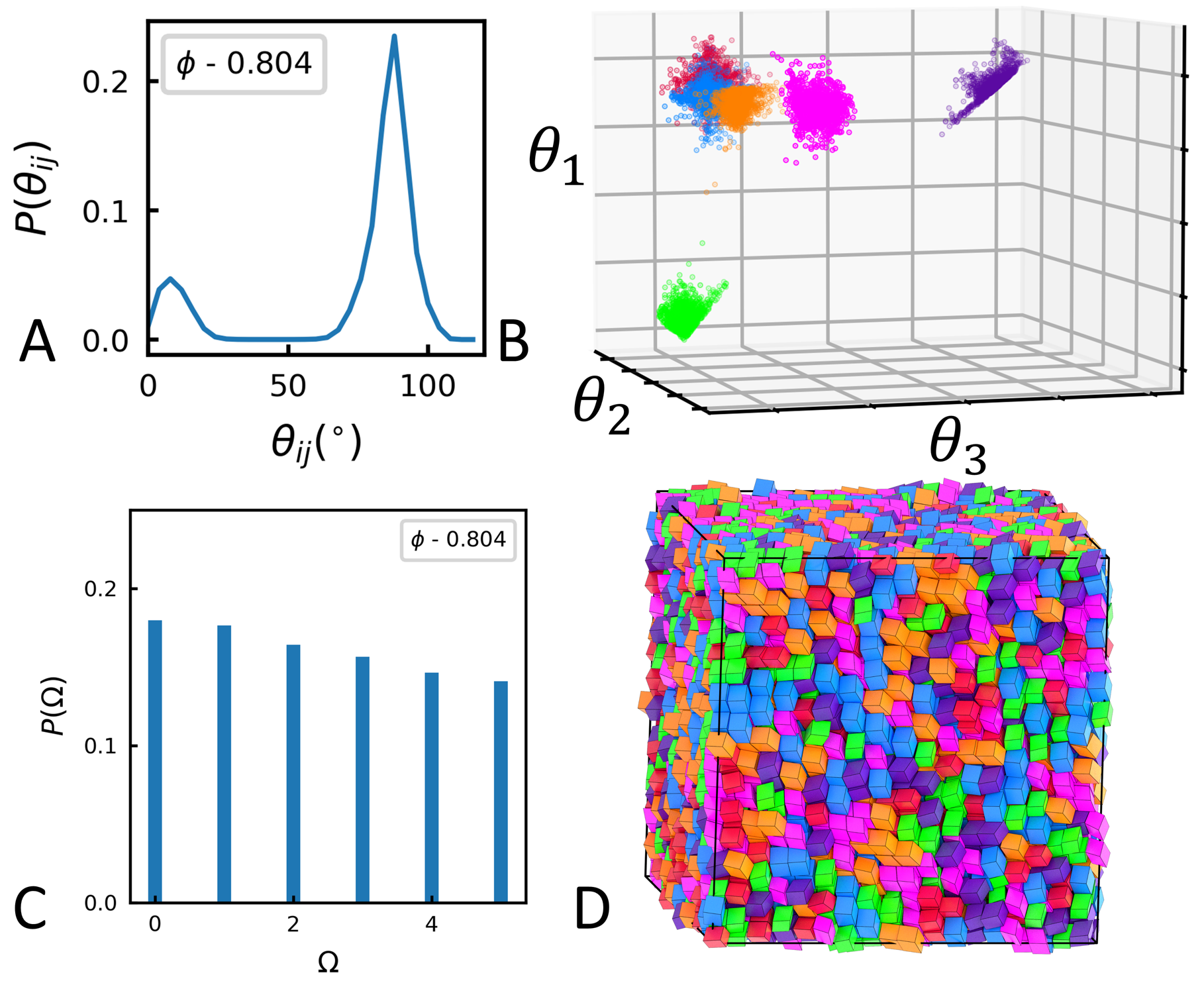}
	\caption{\textbf{Orientational disorder in sheared Cube $20^{\circ}$:} A. Distribution of pairwise angles B. Absolute orientations of all
		particles in the simulation system in a three dimensional space of orientational differences
		from three reference orientations ($\theta_{1}, \theta_{2}, \theta_{3}$) C. Population distribution among these 6 unique orientations D. snapshots of simulation systems with uniquely oriented particles
		rendered in a single color}
	\label{fig:deg20_si}
\end{figure*}

\begin{figure*}
	\centering 
	\includegraphics[scale=1.0]{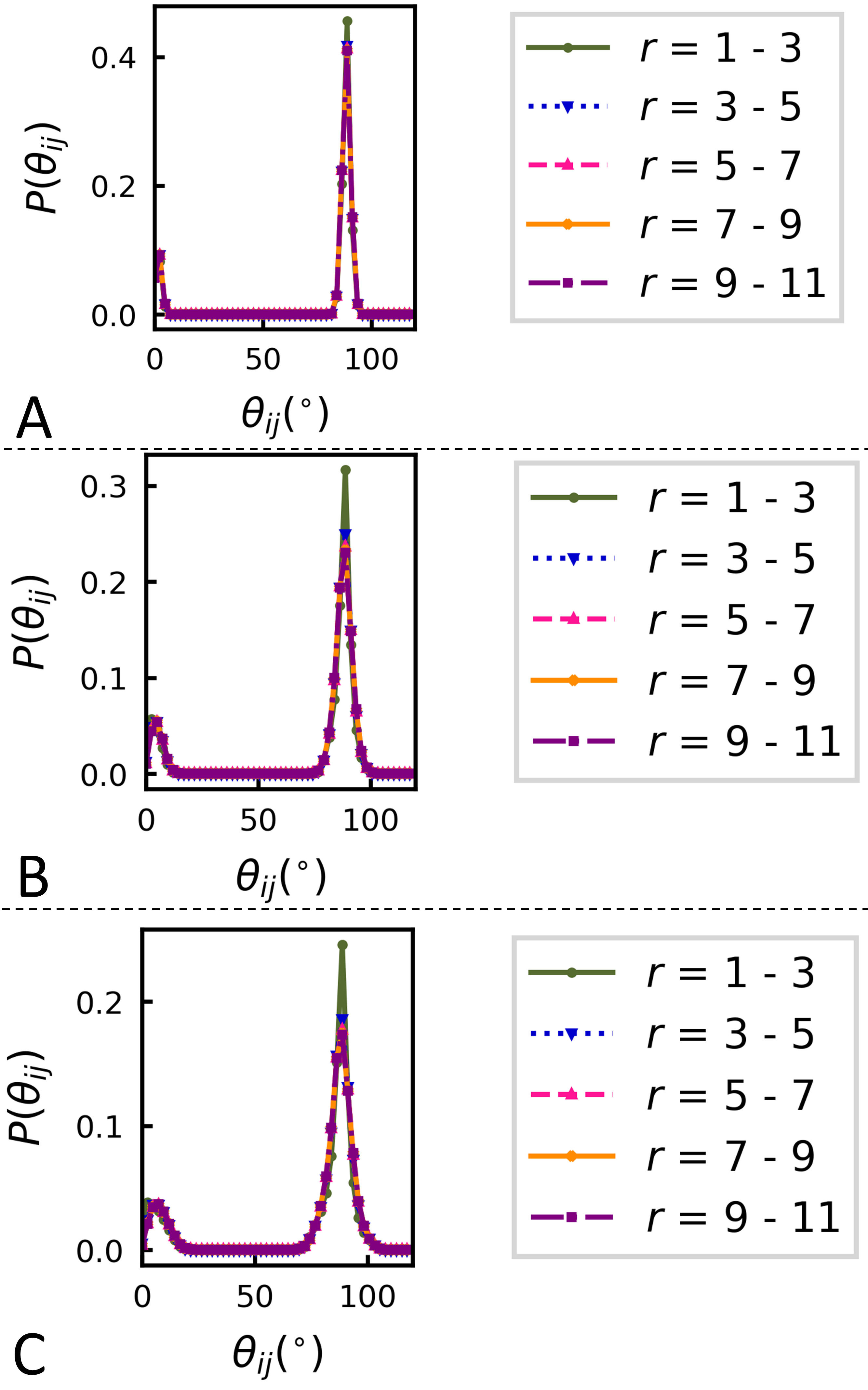}
	\caption{\textbf{Distance invariance of distribution of pairwise orientational differences:} A. Sheared Cube $5^{\circ}$ B. Sheared Cube $10^{\circ}$ C. Sheared Cube $15^{\circ}$ Distribution of pairwise orientational differences of particles at different lengths}
	\label{fig:distance_inv_supplementary}
\end{figure*}

\begin{figure*}
	\centering 
	\includegraphics[scale=1.0]{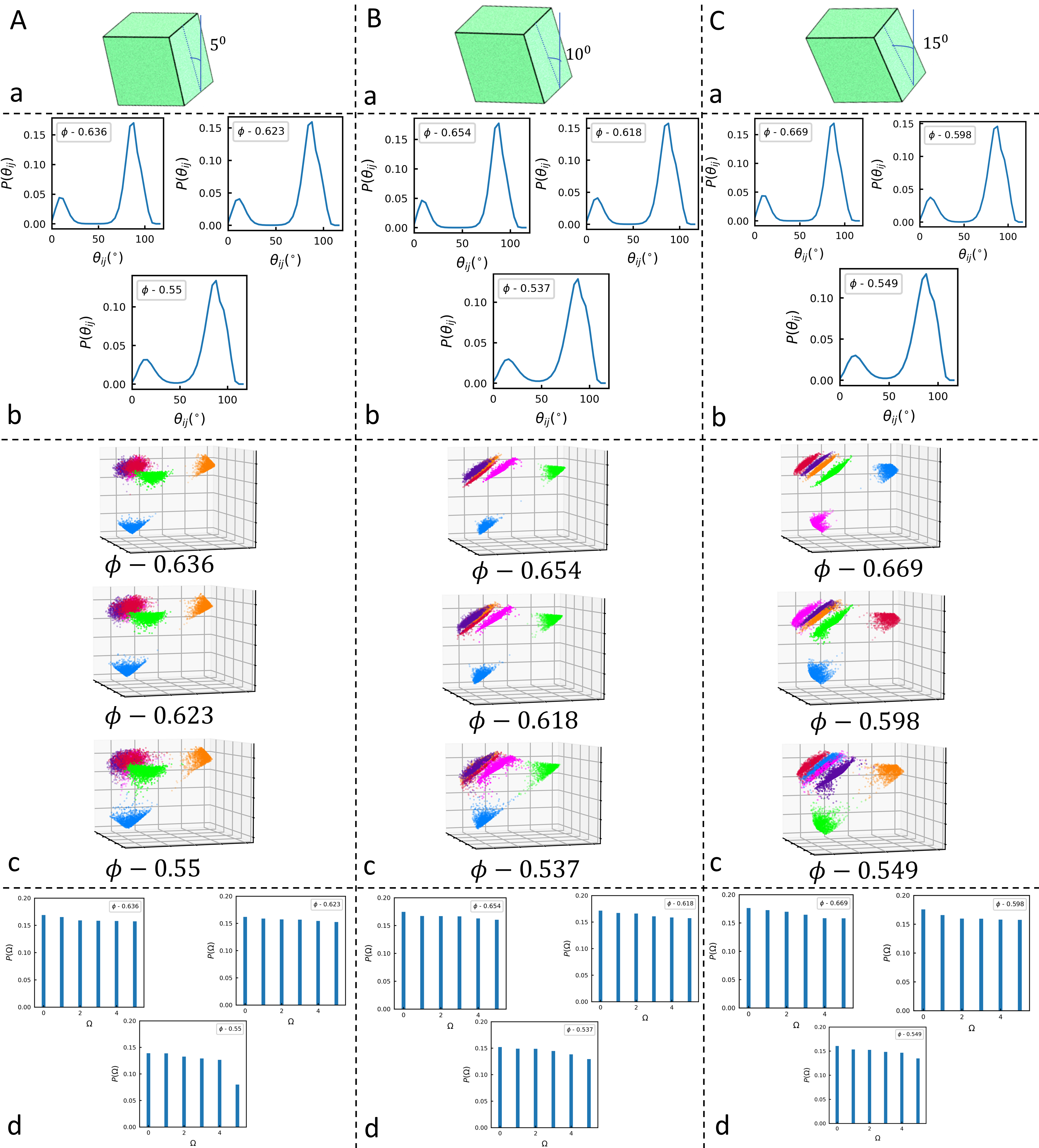}
	\caption{\textbf{Orientational distributions at lower packing fractions:} A. Sheared Cube $5^{\circ}$ B. Sheared Cube $10^{\circ}$ C. Sheared Cube $15^{\circ}$ a. Sheared cube shapes b. Distribution of pairwise orientational differences c. Orientations of all particles in the simulation system in a three dimensional space of orientational differences from three reference orientations from system d. Population distributions among six distinct unique orientations}
	\label{fig:corr_dis_at_lower_pf}
\end{figure*}

\begin{figure*}
	\centering 
	\includegraphics[scale=0.24]{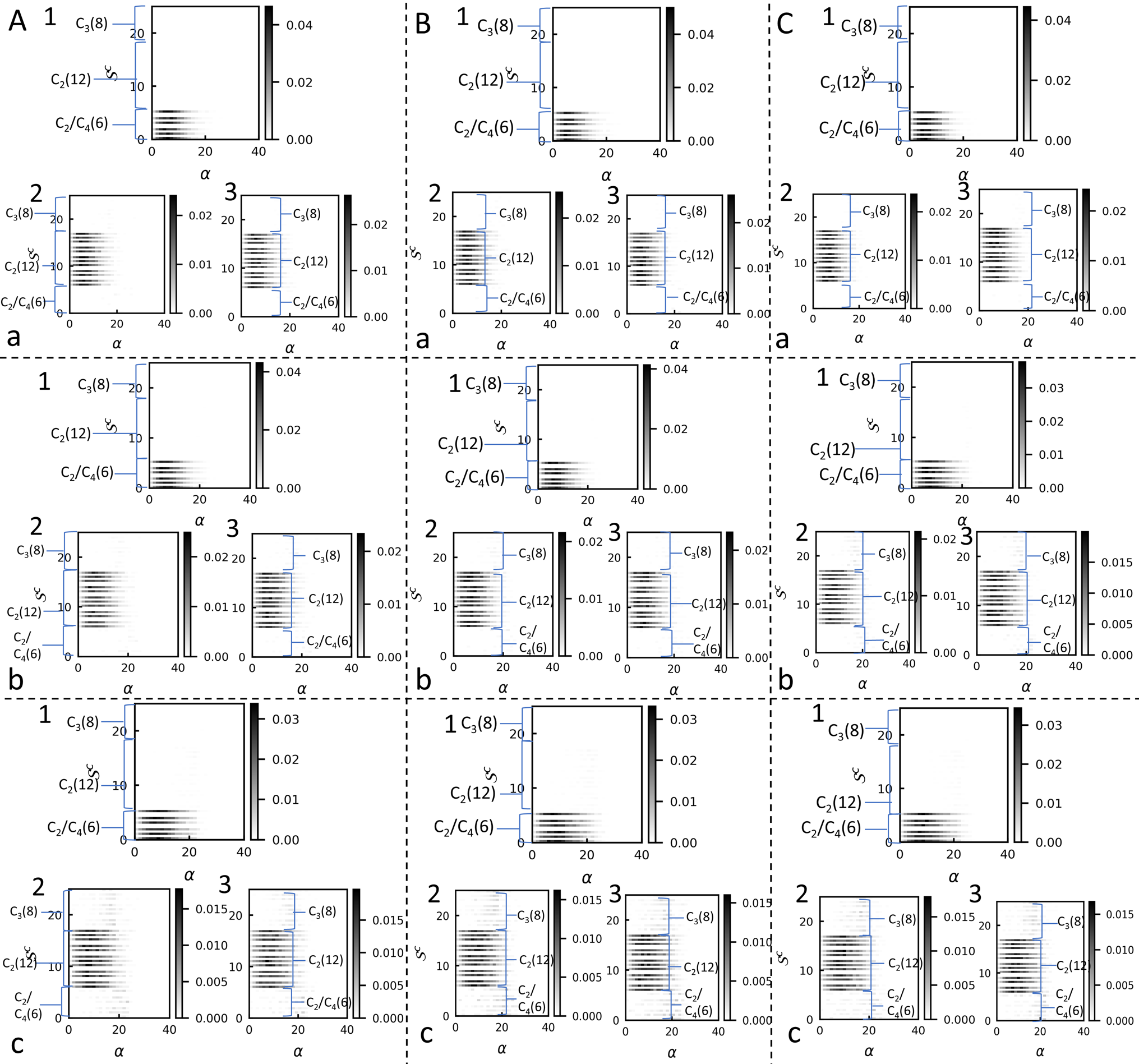}
	\caption{\textbf{Alignments of $\mathcal{S}^{p}_{max} (C_{2})$ axes of the particle with different $\mathcal{S}^{c}$ axes at lower density solids:} A. Sheared Cube $5^{\circ}$ $\mathcal{S}^{p}_{max} (C_{2})$ axes alignments at a. $\phi \sim 0.636$ b. $\phi \sim 0.623$ c $\phi \sim 0.55$ B. Sheared Cube $10^{\circ}$ $\mathcal{S}^{p}_{max} (C_{2})$ axes alignments at a. $\phi \sim 0.654$ b. $\phi \sim 0.618$ c $\phi \sim 0.537$ C. Sheared Cube $15^{\circ}$ $\mathcal{S}^{p}_{max} (C_{2})$ axes alignments at a. $\phi \sim 0.669$ b. $\phi \sim 0.598$ c $\phi \sim 0.549$. For each of the rows for each sheared cubes 1. Alignment of $\mathcal{S}^{p}_{max} (C_{2})$ passing through faces 2. Alignment of $\mathcal{S}^{p}_{max} (C_{2})$ passing through edges 3. Alignment of $\mathcal{S}^{p}_{max} (C_{2})$ passing through edges }
	\label{fig:axis_alignment_at_lower_pf}
\end{figure*}						
						
\end{document}